\documentclass[%
 reprint,
 amsmath,amssymb,
 aps,
 floatfix,
 longbibliography,
]{revtex4-2}

\usepackage{graphicx}
\usepackage{bm}
\usepackage[utf8]{inputenc}
\usepackage[T1]{fontenc}
\usepackage[%
  colorlinks=true,
  urlcolor=blue,
  linkcolor=blue,
  citecolor=blue]{hyperref}
\usepackage[capitalise]{cleveref}
\usepackage{hyperref}

\begin{document}

\title{\texorpdfstring{\textit{Ab
Initio} Study of Magnetic Tunnel Junctions Based on Half-Metallic and
Spin-Gapless Semiconducting Heusler Compounds: Reconfigurable Diode
and \\ Inverse Tunnel-Magnetoresistance Effect}{Ab
Initio Study of Magnetic Tunnel Junctions Based on Half-Metallic and
Spin-Gapless Semiconducting Heusler Compounds: Reconfigurable Diode
and Inverse Tunnel-Magnetoresistance Effect}}
\author{T. Aull} 

\author{E. \c{S}a\c{s}{\i}o\u{g}lu}

\author{N. F. Hinsche}

\author{I. Mertig}
\affiliation{Institute of Physics, Martin Luther University Halle-Wittenberg, D-06120 Halle (Saale), Germany}

\date{\today}

\begin{abstract}
Magnetic tunnel junctions (MTJs) have attracted strong research interest within the last decades due 
to their potential use as nonvolatile memory such as magnetoresistive random access memory as well as for magnetic logic applications. 
Half-metallic magnets (HMMs) have been suggested as ideal electrode materials for MTJs to 
achieve an extremely large tunnel-magnetoresistance (TMR) effect. Despite their high TMR ratios,
MTJs based on HMMs do not exhibit current rectification, i.e., a diode effect, which was 
achieved in a  
magnetic tunnel junction concept [ACS Appl. Electron. Mater. \textbf{1}, 1552–9 (2019)]
based on HMMs and type-II spin-gapless semiconductors (SGSs). The proposed concept has recently been  experimentally demonstrated using Heusler compounds. In the present work, we investigate 
from first-principles MTJs based on type-II SGS and HMM quaternary Heusler compounds FeVTaAl, 
FeVTiSi, MnVTiAl, and CoVTiSb. Our \textit{ab initio} quantum transport calculations based on a
nonequilibrium Green's function method have demonstrated that the MTJs under consideration exhibit 
current rectification with relatively high on:off ratios. We show that, in contrast to conventional 
semiconductor diodes, the rectification bias voltage window (or breakdown voltage) of the MTJs is 
limited by the spin gap of the HMM and SGS Heusler compounds. A unique feature of the present MTJs is that the diode effect can be configured dynamically, 
i.e., depending on the relative orientation of the magnetization of the electrodes, the MTJ allows the 
electrical current to pass either in one or the other direction, which leads to an inverse TMR effect. 
The combination of nonvolatility, reconfigurable diode functionality, tunable rectification voltage 
window, and high Curie temperature of the electrode materials makes the proposed MTJs very promising 
for room-temperature spintronic applications and opens  
ways to magnetic memory and logic concepts as well as logic-in-memory computing.
\end{abstract}


\maketitle

\section{INTRODUCTION}

The current computing technology is based on the von Neumann architecture~\cite{von1993first}, in which the 
central processing unit and the memory are connected via a shared bus system causing the memory bandwidth 
bottleneck and high power consumption. It was demonstrated that for many computing tasks, the majority 
of the energy and time is needed to transfer data between the memory and the CPU, rather than the information 
processing itself~\cite{horowitz2014computing,li2015learnable}. To tackle the bandwidth bottleneck in today's 
microprocessors,  
information processing concepts such as logic-in-memory computing are receiving substantial 
interest~\cite{linn2012beyond,you2014exploiting,gao2015implementation,zhou2015boolean,wang2017functionally,kim2019single}.
The logic-in-memory computing architecture requires nonvolatile memory elements. Among the emerging nonvolatile
memory technologies, the magnetoresistive random access memory (MRAM) is the most promising candidate due to its almost 
infinite endurance. The MRAM combines relatively high access speeds with nonvolatility. In particular, spin-transfer
torque (STT) MRAM and spin-orbit torque (SOT) MRAM emerged as promising candidates to replace the L3  
and L2 caches~\cite{xu2011design,prenat2016ultra} of modern microprocessors.

\begin{figure}[!t]
\centering
\includegraphics[width=0.45\textwidth]{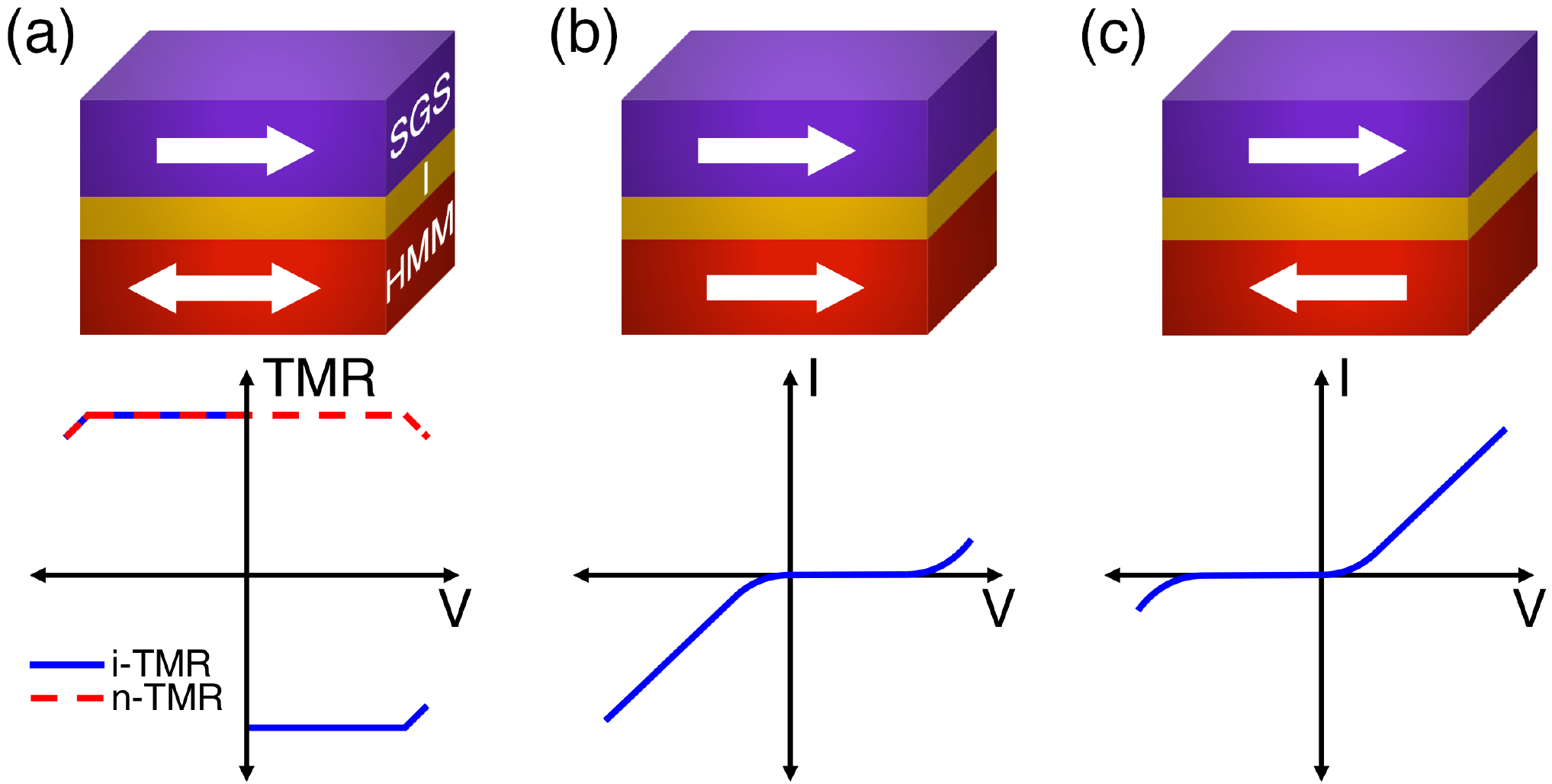}
\caption{(a) Top: schematic representation of the magnetic tunnel junction based on a half-metallic 
magnet and a spin gapless semiconductor. Bottom: Dependency of the TMR effect on the bias voltage in MTJs. 
The inverse (i-)TMR effect is illustrated by a blue line while the normal (n-)TMR effect is represented by a 
red dashed line. (b),(c) The same as (a) for the parallel and antiparallel orientations of the magnetization 
directions of the electrodes as well as the corresponding current-voltage ($I$-$V$) characteristics. The white arrows 
indicate the magnetization direction of the electrodes.}
\label{fig1}
\end{figure}

In conventional magnetic tunnel junctions (MTJs), a nonmagnetic insulator of a few nanometer thickness is sandwiched 
between two ferromagnetic electrodes~\cite{yuasa2004giant,parkin2004giant}. Thus, the electronic transport is 
spin dependent and mainly determined by quantum tunneling. For this reason, the tunnel-magnetoresistance (TMR) 
ratio and the conductance are very important quantities of MTJs~\cite{ebke2006large,kou2006temperature,reiss2006co,krzysteczko2009current}. 
The resistance of such devices differs in two configurations, when the magnetization of the left and right electrodes is 
parallel oriented and when the orientation is switched to antiparallel, resulting in the TMR effect. When no bias 
voltage is applied, the TMR ratio is defined as $\text{TMR} = \left(G_{\uparrow \uparrow}-G_{\uparrow \downarrow}\right)/
(G_{\uparrow \downarrow} + G_{\uparrow \uparrow})$, where $G_{\uparrow \uparrow}$ ($G_{\uparrow \downarrow}$) 
denotes the conductance in the parallel (antiparallel) configuration of the electrodes. For finite biases, the 
TMR expression becomes $\text{TMR} = \left(I_{\uparrow \uparrow}-I_{\uparrow \downarrow}\right)/(I_{\uparrow \downarrow}
+ I_{\uparrow \uparrow})$, where $I_{\uparrow \uparrow}$ ($I_{\uparrow \downarrow}$) is the tunnel current through 
the device in the parallel (antiparallel) orientation of the magnetization of the electrodes. It is worth noting that the tunnel 
barrier material, as well as the thickness of the tunnel barrier, and the applied bias voltage can influence the TMR 
effect~\cite{mathon2001theory,parkin2004giant,ikeda2008tunnel}. Another factor that can affect the sign and the value 
of the TMR ratio is a structural asymmetry in the junctions. Heiliger \textit{et al.}~\cite{heiliger2005influence,heiliger2006interface} proposed that, independent 
of the applied bias voltage, in asymmetric junctions the value of $I_{\uparrow \downarrow}$ exceeds the value 
of $I_{\uparrow \uparrow}$ and, as a consequence, leads to a negative TMR ratio. 
The dependency of the TMR ratio on the applied bias voltage for both the normal and the inverse TMR effect is 
schematically illustrated in Fig.~\hyperref[fig1]{1\,(a)}.

MTJs played a significant role in spintronics development as they are suitable for several applications 
ranging from read-head sensors to nonvolatile memory devices such as STT MRAM and SOT MRAM and from 
nonvolatile logic concepts to logic-in-memory computing~\cite{jain2017computing,kingra2020slim,de2021stt}. 
Magnetic logic promises nonvolatile, low-power computing and up to now, several different approaches 
have been proposed such as the quantum cellular automata~\cite{cowburn2000room,amlani1999digital}, 
domain-wall logic~\cite{allwood2005magnetic,tsoi2003magnetic}, MTJ logic~\cite{richter2002nonvolatile,ney2003programmable,thomas2006inverted}, 
etc. The latter is of particular interest because it opens the way to logic-in-memory computing, i.e., 
storing and processing the data within the same chip and thus providing an opportunity to explore  
computing architectures beyond the classical von Neumann architecture~\cite{dery2007spin,behin2010proposal}. 
MTJ-based magnetic logic proposals can be divided into three categories: i) external field-driven MTJ logic, 
ii) spin Hall effect driven MTJ logic, and iii) logic based on magnetic tunnel diodes and magnetic tunnel 
transistors.

In the first category, the logic gates are built from MTJs, which are arranged in a bridge-type 
configuration and the logic inputs are provided by external wires, which creates a magnetic field 
that switches the magnetization direction of one electrode in the MTJ. In this way, all logic gates 
can be realized with few MTJs~\cite{richter2002nonvolatile,ney2003programmable,friedman2014complementary}. 
The utilization of an inverse TMR effect can even further reduce the number of MTJs in logic 
gates~\cite{thomas2006inverted,isogami2014electronic}. However, the drawback of this approach is 
that it is not scalable due to input wires and their routing near the MTJs. In the second category, 
the logic gates are based on a  
four-terminal spin Hall effect driven MTJ with fully electrically-separated 
write and read paths~\cite{bromberg2012novel,fukami2009low,kang2015complementary}. These four-terminal 
MTJ devices can overcome the challenges of operation gain and direct cascading in current spintronic 
logic circuits. Moreover, simulations have indicated that correct logic fan-out operation can be achieved 
with voltage below 150\,mV, which is promising for low-power computing~\cite{kang2015complementary}. 
Note that in both approaches the logic operation gain (i.e., output voltage margin) depends mainly 
on the TMR ratio of the MTJs. While in the third category, a MTJ possesses, in addition to the TMR effect 
(memory), a current rectification (diode effect) functionality. Such MTJs also constitute the basic 
building blocks of the three-terminal magnetic tunnel transistors for logic applications. The TMR effect 
and current rectification have been observed for single barrier asymmetric MTJs as well as for double 
barrier MTJs with tunnel barriers of different transparencies~\cite{de2006low,chshiev2002magnetic,tiusan2001quantum,iovan2008spin}.
Although in initial studies of magnetic tunnel transistors low magnetocurrent ratios and transfer 
rates $\alpha$ are reported~\cite{van2002spin,sato2001spin,rodary2005development,gobbi2012c60,lee2005large}, 
in recent experiments of fully epitaxial magnetic tunnel transistors a large magnetocurrent ratio and 
transfer rate $\alpha$ is detected~\cite{nagahama2010hot}. Besides being the basic building blocks 
of the three-terminal magnetic tunnel transistors, the MTJ possessing the diode effect is of particular 
interest for high-density three-dimensional (3D) cross-point STT MRAM applications as it eliminates the need for an additional 
selection device~\cite{iovan2008spin}, i.e., a MOSFET transistor or a $p-n$ diode~\cite{aluguri2016notice,burr2014access,peng2017cross}.

In contrast to MTJ-based logic proposals, in the first and second category as well as other concepts like 
spin-orbit torque logic~\cite{nikonov2011proposal} not mentioned above (for a detailed discussion, we refer the reader to 
Refs.~\onlinecite{nikonov2015benchmarking,pan2017expanded}, which report a benchmarking 
of beyond-CMOS devices including various spintronic logic concepts), magnetic tunnel diodes and transistors 
can operate extremely high frequencies, i.e., in the terahertz regime, making them ideal candidates for high-speed 
electronic and spintronic applications. However, despite terahertz operation frequencies, conventional magnetic 
tunnel diodes and transistors come with fundamental issues such as low on:off current ratios and less asymmetric 
current-voltage characteristics in diodes and base-collector leakage currents in transistors, which might 
lead to high power dissipation. In Ref.~\onlinecite{sasioglu2019proposal}, we proposed a magnetic tunnel diode and transistor concept, which overcomes the limitations of conventional magnetic tunnel devices and 
provides additional unique functionalities like reconfigurability, which was recently experimentally demonstrated~\cite{maji2022demonstration}. 
The concept is based on spin-gapless semiconductors (SGSs)~\cite{wang2008proposal} and half-metallic magnets 
(HMMs)~\cite{de1983new}. The two-terminal magnetic tunnel diode (or MTJ) is comprised of a SGS electrode and 
a HMM electrode separated by a thin insulating tunnel barrier. A schematic representation of the structure of 
the reconfigurable magnetic tunnel diode is shown in Fig.~\hyperref[fig1]{1\,(b)} and \hyperref[fig1]{1\,(c)}.
Depending on the relative orientation of the magnetization of the electrodes the MTJ allows the electrical 
current to pass either in one or the other direction.

The aim of the present paper is a computational design of MTJs based on HMM and SGS quaternary Heusler 
compounds for room-temperature device applications. Heusler compounds offer a unique platform to realize 
MTJs as these materials possess very high Curie temperatures (above room-temperature) as well as 
HMM and SGS behavior within the same family~\cite{ozdougan2013slater,gao2019high,aull2019ab}. 
To this end, the selection of the HMM and SGS electrode materials from the quaternary Heusler family 
for the design of MTJs is based on our recent study in Ref.~\onlinecite{aull2019ab}. We stick to SGS 
FeVTaAl and FeVTiSi compounds due to their large energy gaps in opposite spin channels around the Fermi 
level~\cite{aull2019ab}, MgO as the tunnel barrier due to the lattice matching, and for the HMMs, although 
we have a large variety of choice, we choose MnVTaAl and CoFeVSb since both materials exhibit nearly 
symmetric spin gaps above and below $E_F$ and possess similar lattice constants to MgO. \textit{Ab initio} 
quantum transport calculations based on the nonequilibrium Green’s function (NEGF) method have demonstrated 
that the MTJs based on HMM and SGS Heusler compounds exhibit, in addition to the inverse TMR effect, current 
rectification, i.e, diode effect, which can be dynamically configured. We show that in contrast to 
semiconductor diodes ($p-n$ diode or Schottky diode), the rectification voltage window (or breakdown voltage) 
of these MTJs is limited by the spin gap of HMM and SGS Heusler compounds.
The calculated zero temperature on:off current ratios vary between 
$10^2-10^7$, being lowest for the FeVTiSi/MgO/CoFeVSb MTJ,
which can be attributed to the overlap of the conduction and valence bands of opposite 
spin channels around the Fermi level. The combination of nonvolatility and the dynamically reconfigurable diode 
effect as well as the very high Curie temperature of quaternary Heusler compounds makes the proposed MTJs very 
promising for room-temperature spintronic memory and logic applications. The rest of the paper is organized as 
follows. In \cref{sec:junction}, we discuss the $I$-$V$ characteristics of the MTJ concept by using 
the spin dependent energy-band diagrams. In \cref{sec:computDetail}, we present the computational details of 
our study. Our computational results are presented and discussed in \cref{sec:results}, and finally, in 
\cref{sec:concl}, we give our summary and outlook.

\section{\label{sec:junction}HMM/I/SGS MAGNETIC TUNNEL JUNCTIONS}

\begin{figure}[!t]
\vspace{-1em}
\centering
\includegraphics[width=0.48\textwidth]{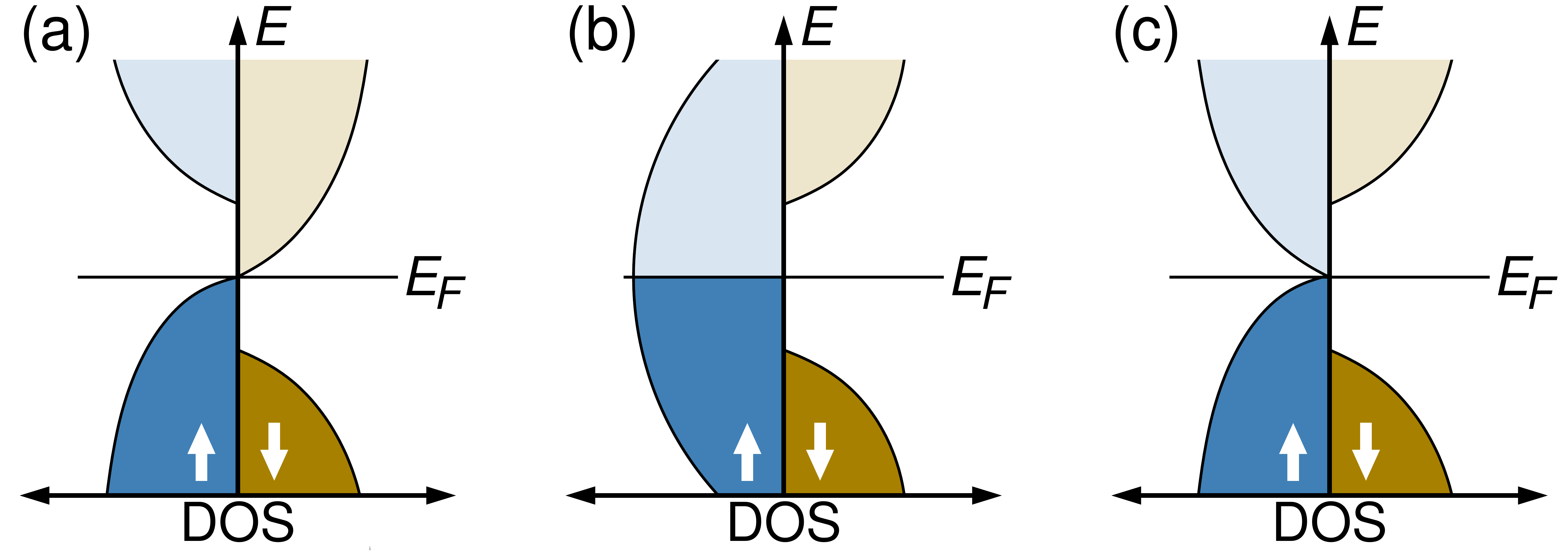}
\caption{Schematic representation of the density of states for a (a) type-II spin-gapless semiconductor, 
(b) half-metallic magnet, and (c) type-I spin-gapless semiconductor.}
\label{fig2}
\end{figure}

In Fig.~\hyperref[fig1]{1\,(b)} and \hyperref[fig1]{1\,(c)}, we schematically show a MTJ based on a HMM 
and a SGS in the parallel and antiparallel configurations of the electrodes, respectively, together with 
the corresponding $I$-$V$ curves. HMMs have been used as electrode materials for MTJs to achieve extremely large 
TMR effects. Despite their large TMR ratios, the MTJs based on HMMs do not present current rectification, i.e., 
a diode effect. In Ref.~\onlinecite{sasioglu2019proposal}, it was proposed that replacing one of the HMM electrodes 
with a SGS material in a MTJ gives rise to additional functionalities, i.e., current rectification, inverse TMR effect, 
and reconfigurability of the MTJ. Such a MTJ is also called a reconfigurable magnetic tunnel diode~\cite{sasioglu2019proposal}.
Besides the HMM, the SGS material is the key component of this MTJ. SGSs have been proposed by Wang in 2008 
as a theoretical concept~\cite{wang2008proposal}. By employing first-principles calculations Wang demonstrated 
that doping PbPdO$_2$ with Co atoms results in a new class of materials: the SGSs~\cite{wang2008proposal,wang2009colossal}. 
Since then, different classes of materials have been predicted to present SGS behavior of various types, i.e., 
from type-I to type-IV SGSs~\cite{wang2008proposal,xu2013new,skaftouros2013search,ozdougan2013slater,galanakis2016spin,gao2019high,aull2019ab}
and some of the predicted SGSs have been experimentally realized~\cite{ouardi2013realization}. Since type-II SGSs 
are the key component of the reconfigurable MTJ, in Fig.~\ref{fig2} we present the schematic density of states (DOS) 
of a type-II SGS together with a conventional HMM as well as a type-I SGS, which can also be used as a replacement 
of the HMM in a MTJ. As seen in Fig.~\hyperref[fig2]{2\,(a)} the type-II SGS possesses a unique electronic band structure, 
i.e., it presents a finite gap below and above the Fermi level $E_F$ in different spin channels, while the valence 
and conduction bands of different spin channels touch at $E_F$. On the other hand, in HMMs, the majority-spin channel 
behaves like in normal metals, but the minority-spin channel exhibits a gap around the Fermi level like in a semiconductor 
or insulator. The DOS of type-I SGSs is similar to HMMs [see Fig.~\hyperref[fig2]{2\,(b)} and \hyperref[fig2]{2\,(c)}]. 
The minority-spin channel looks the same while in the majority-spin channel a zero-width gap appears at the Fermi level 
since the conduction- and valence-band edges touch at $E_F$.

\begin{figure*}
\centering
\includegraphics[width=0.98\textwidth]{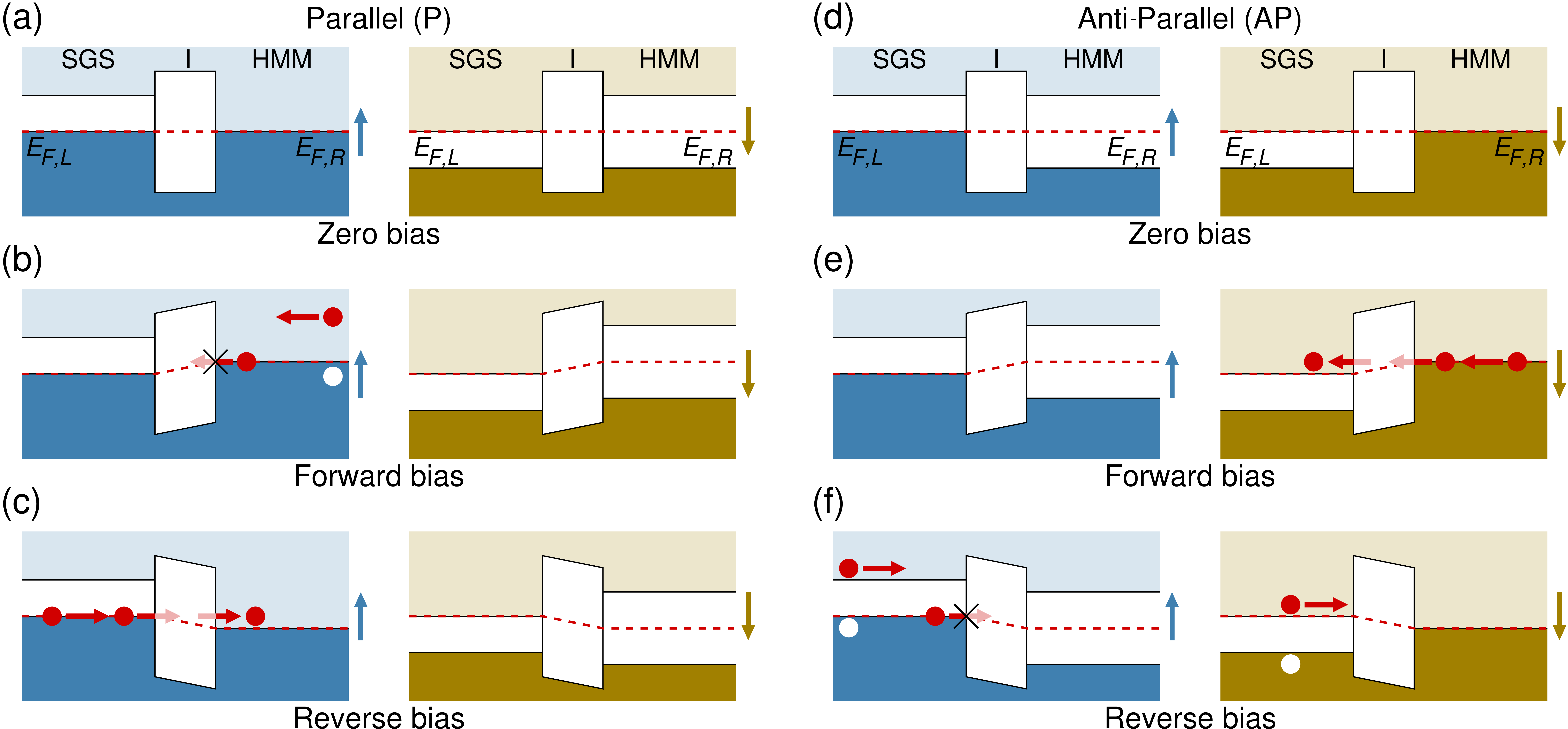}
\caption{Schematic representation of the spin-resolved energy-band diagram for the SGS/I/HMM MTJ
for parallel (P) orientation of the magnetization directions of the electrodes (a) for zero bias, (b) under
forward bias, and (c) under reverse bias. The electrons (holes) and the Fermi energy are denoted by
red (white) spheres and a dashed line, respectively, and the tunneling process is illustrated by partly
shaded red arrows. Panels (d)–(f) represent the same as (a)–(c) for the antiparallel (AP) orientation of
the magnetization directions of the electrodes 
[see Figs.~\hyperref[fig1]{1\,(b)} and \hyperref[fig1]{1\,(c)}].}
\label{fig3}
\end{figure*}

The operation principle of the reconfigurable MTJ is extensively discussed in Ref.~\onlinecite{sasioglu2019proposal} 
and hence here we present a short overview of the concept by using the spin-resolved energy-band diagram shown in 
Fig.~\ref{fig3}. The spin-resolved energy-band diagram is based on the schematic densities of states provided in Fig.~\hyperref[fig2]{2\,(a)} 
and \hyperref[fig2]{2\,(b)}, i.e., the type-II SGS material possesses a gap in the minority-spin (majority-spin) channel 
below (above) the Fermi level while the HMM exhibits a gap in the minority-spin channel around the Fermi energy. We 
further assume that the type-II SGS electrode, the tunnel barrier, and the half-metallic material have the same work 
function and equal Fermi levels and therefore we do not consider charge transfer at the interfaces. However, real materials, 
as will be discussed in \cref{sec:results}, possess different work functions and so there occurs charge transfer between 
one material and the other at the interface, which might cause a band bending in the SGS electrode. Moreover, due to 
interactions at the interface, the junction materials might not conserve the SGS or HMM characteristics close to the interface 
and thus the band diagram will not be as sharp as presented in \cref{fig3}.

The $I$-$V$ characteristics of the SGS/I/HMM junction illustrated in Figs.~\hyperref[fig1]{1\,(b)} and \hyperref[fig1]{1\,(c)} 
can be qualitatively explained by Bardeen’s approach for tunneling~\cite{bardeen1961tunnelling,meservey1994spin}.
For a simple tunnel barrier, the tunnel current $I(V)$ is given by the expression 
\begin{gather*}
    I(V) \sim \sum_{\sigma}\int_{-\infty}^{+\infty} 
    \rho_{\textrm{HMM}}^{\sigma}(E+eV) \, \rho_{\textrm{SGS}}^{\sigma}(E) \, |T(V)|^2 \\
    \times [f(E)-f(E+eV)] \, \mathrm{d}E,
\end{gather*}
where $\rho_{\textrm{SGS}}^{\sigma}(E)$ and $\rho_{\textrm{HMM}}^{\sigma}(E+eV)$ denote the density of states of the SGS and HMM electrodes 
with spin $\sigma $, $f(E)$ is the Fermi distribution function and $T(V)$ is the transmission probability, which is 
proportional to $e^{-d\sqrt{\phi-V}}$, where $d$ is the thickness of the tunnel barrier and $\phi$ is the barrier height.
As shown in Fig.~\hyperref[fig3]{3\,(b)}, when the magnetization directions of the electrodes are aligned parallel and 
a positive bias voltage (forward bias) is applied to the SGS electrode, electrons in the occupied majority-spin valence 
band of the HMM electrode cannot tunnel through the insulating barrier into the SGS electrode because there are no 
available states above the Fermi energy in the majority-spin channel of the SGS electrode unless a certain bias voltage 
is reached. For minority-spin electrons, the HMM electrode behaves like an insulator and thus no electron transport
takes place. For a negative bias voltage (reverse bias), electrons in the majority-spin channel in the SGS material can 
tunnel into the unoccupied states of the HMM as shown in Fig.~\hyperref[fig3]{3\,(c)}. In the minority-spin channel, 
neither in the SGS electrode nor in the HMM electrode are states available that can contribute to a current. Thus, the 
tunneling current through the MTJ is 100\,\% spin polarized. A similar discussion holds for the antiparallel orientation 
of the magnetization direction of the SGS and HMM electrodes, for which the corresponding energy-band diagram is presented 
in Figs.~\hyperref[fig3]{3\,(d)}-\hyperref[fig3]{(f)}. Note that in the schematic representation of the $I$-$V$ characteristics 
of the MTJ [see Figs.~\hyperref[fig1]{1\,(b)} and \hyperref[fig1]{1\,(c)}], we use the standard definition of current for 
semiconductor devices, i.e., the current direction is opposite to the electron motion direction, while in 
Ref.~\onlinecite{sasioglu2019proposal}, the same direction is taken for the current and electron motion. This is why 
the $I$-$V$ characteristics are different in Ref.~\onlinecite{sasioglu2019proposal}.

\begin{table*}
\caption{\label{tab1}
Material compositions of the considered MTJs, lattice constants $a_0$, $c/a$ ratio, sublattice and total magnetic moments, 
work function ($\Phi$), Curie temperatures $T_C$ of the cubic phase, and the electronic ground state. 
All $T_C$ values are taken from Ref.~\onlinecite{aull2019ab}.}
\begin{ruledtabular}
\begin{tabular}{@{}*{1}l*{11}{c}@{}}
SGS/MgO/HMM junction & MgO-interface & Compound & $a_0$ & $c/a$ & m$_X$ & m$_{X'}$ & m$_Y$ & m$_\text{total}$ & $\Phi$ & $T_C$ & Ground state \\
 & & & ({\AA}) & & ($\mu_B$) & ($\mu_B$) & ($\mu_B$) & ($\mu_B$) & (eV) & (K) \\
\hline
FeVTaAl/MgO/MnVTiAl & FeV-MnV & FeVTaAl & 6.10 & 1.00 & 0.85 & 2.38 & -0.19 & 3.00 & 3.75 & 681 & SGS \\
  &  & MgO & 6.10 & 0.98 &  &  &  &  & 4.53 &  & I\\
                    &             & MnVTiAl & 6.10 & 1.01 & -2.42 & 2.61 & 0.86 & 1.00 & 3.59 & 963 & HMM \\
FeVTiSi/MgO/CoFeVSb & FeV-CoFe & FeVTiSi & 5.91 & 1.00 & 0.57 & 2.33 & 0.10 & 3.00 & 3.52 & 464 & SGS \\
  &  & MgO & 5.91 & 1.04 &  &  &  &  & 4.55 &  & I\\
  &  & CoFeVSb & 5.91 & 1.12 & 1.08 & 1.20 & 0.78 & 3.00 & 4.10 & 308 & HMM 
\end{tabular}
\end{ruledtabular}
\end{table*}

The presence of the reconfigurable diode effect in MTJs based on SGSs and HMMs leads to an inverse TMR effect rather 
than a normal TMR effect as in most of the conventional MTJs. The voltage dependence of the TMR presented in 
Fig.~\hyperref[fig1]{1\,(a)} can be explained on the basis of the $I$-$V$ characteristics discussed above. For 
a positive (forward) bias voltage, $I_{\uparrow \downarrow}$ will take a finite value while $I_{\uparrow \uparrow}$ 
is equal to zero. For a negative (reverse) bias voltage, the situation is exactly the opposite. Thus, for 
forward bias, the TMR ratio will take the value $-100$\,\% in a bias voltage window, which is set by the band gap 
of the SGS and HMM electrodes. Similarly, under a reverse bias, the TMR ratio will be normalized to $+100$\,\%. Note 
that we use here a different definition of the TMR ratio compared to the Julli{\`e}re model~\cite{julliere1975tunneling}.

Up to now, the discussion of the $I$-$V$ curves and voltage dependence of the TMR effect in SGS/I/HHM MTJs
was based on the schematic energy-band diagram at zero temperature and perfect SGS behavior of the electrode material.
However, at finite temperatures, thermally excited electrons (non-spin-flip processes) can be transmitted 
from one electrode to the other in the off state and thus cause a leakage current [see Figs.~\hyperref[fig3]{3\,(b)} 
and \hyperref[fig3]{3\,(f)}]. This reduces the on:off and TMR ratios. Nevertheless, such processes can 
be significantly reduced by increasing the band gap of the SGS and HMM materials as the Fermi-Dirac distribution 
function decays exponentially with increasing energy. Besides thermally excited non-spin-flip electrons, spin-flip
processes stemming from spin-orbit coupling and electron-magnon interaction can also reduce the on:off ratio as
well as the TMR effect~\cite{bratkovsky1998assisted,inoue1999effects,schmalhorst2005inelastic}.

\section{\label{sec:computDetail}COMPUTATIONAL DETAILS}

Our \textit{ab initio} study of the SGS/MgO/HMM MTJs is based on spin polarized density functional 
theory (DFT) using the \textsc{QuantumATK} software package (version S-2021.06)~\cite{QuantumATK,smidstrup2019an}.
We use linear combinations of atomic orbitals as the basis-set together with norm-conserving PseudoDojo 
pseudopotentials~\cite{QuantumATKPseudoDojo} with the Perdew-Burke-Ernzerhof (PBE) parametrization of the 
exchange-correlation functional~\cite{perdew1996generalized}. For the determination of the ground-state 
properties, we use a $15 \times 15 \times 15$ Monkhorst-Pack $\mathbf{k}$-point grid and as density mesh cutoff 
for the separation of core and valence electrons 145 Hartree. Since the PBE parametrization is well-known to underestimate 
band gaps~\cite{perdew1999accurate,perdew2017understanding,borlido2019large}, we use the DFT-1/2 
method~\cite{ferreira2008approximation,ferreira2011slater} as implemented in the \textsc{QuantumATK} 
package to correct the band gap in the calculations of the transmission spectra. The changes in the SGS and 
HMM band structure by employing the DFT-1/2 method are negligible. For the structural optimization, all forces 
converge to at least $0.01$\,eV/{\AA} and self-consistency was achieved when the energies between two steps of 
the SCF cycle differ less than $10^{-4}$\,eV. For the transport calculations, we employ the NEGF approach combined with the DFT method using an $11 \times 11 \times 115$ $\mathbf{k}$-point mesh. For all calculations, the smearing is set to 26\,meV.
To calculate the $I$-$V$ characteristics, \textsc{QuantumATK} applies the Landauer-B{\"u}ttiker approach~\cite{Landauer-Buettiker}, 
where $I(V) = (e/h) \sum_{\sigma}\int \, T^{\sigma}(E,V)\left[f_{L}(E,V)-f_{R}(E,V)\right] \mathrm{d}E $,
where $f_L(E,V)$ and $f_R(E,V)$ represent the Fermi-Dirac distributions of the left and right electrodes, 
respectively. Furthermore, the transmission coefficient $T^\sigma (E,V)$ depends on the spin $\sigma$ of the 
electrons, the applied bias voltage $V$, and the energy $E$. For the calculation of $T^\sigma (E,V)$, we choose 
a dense $100 \times 100$ $\mathbf{k}$-point mesh. Moreover, the self-consistent $I$-$V$ calculations are compared 
with a zero-bias linear response approach. It is worth noting that for the linear response approach, the transmission spectrum is treated as bias independent, and thus this approximation is not valid for large biases and therefore just allows for a qualitative description of the $I$-$V$ characteristics.

\section{\label{sec:results}RESULTS AND DISCUSSION}

\begin{figure*}
\centering
\includegraphics[width=0.95\textwidth]{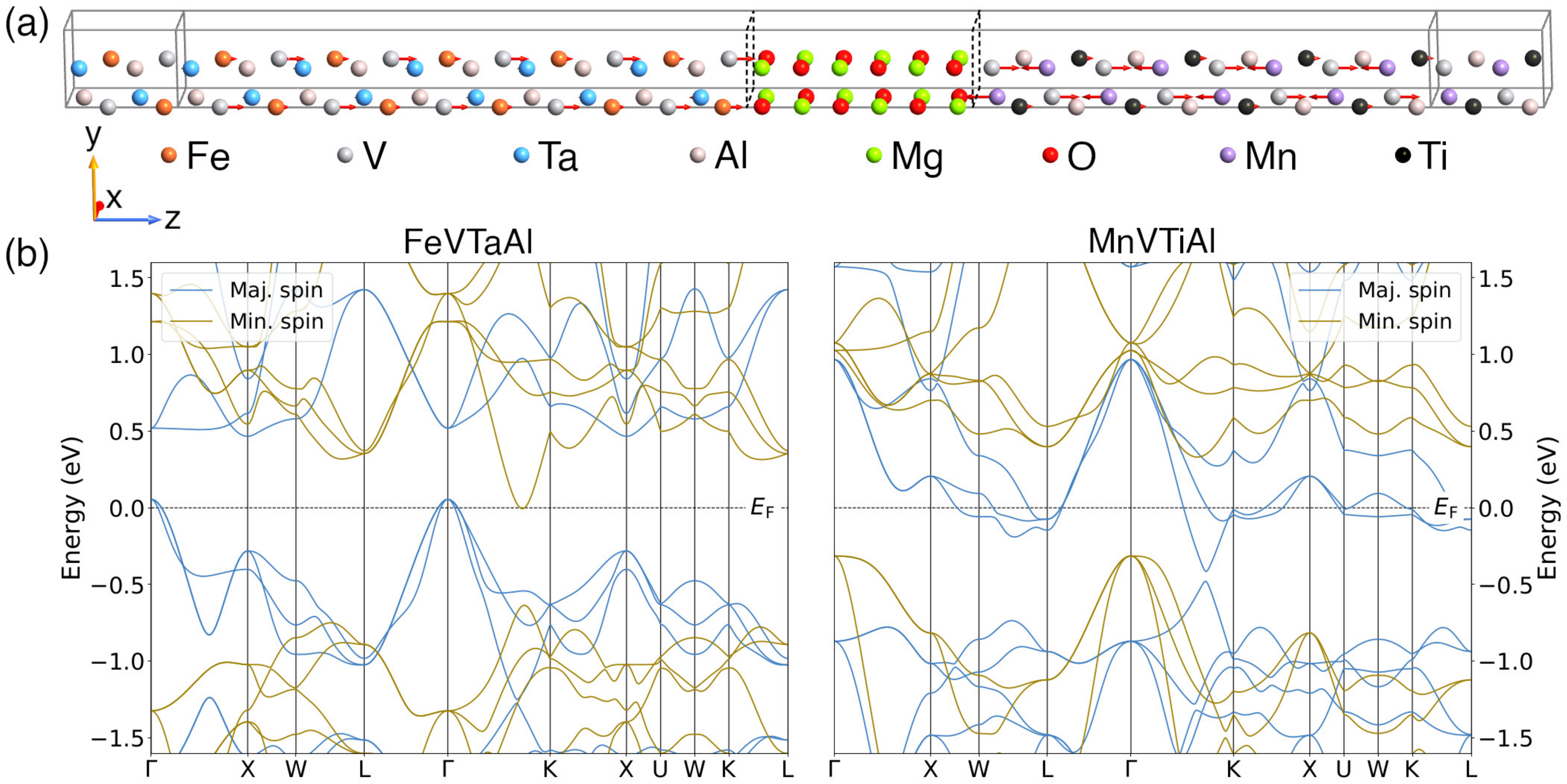}
\caption{(a) The atomic structure of the FeVTaAl/MgO/MnVTiAl tunnel junction. The system is 
periodic in the x-y plane orthogonal to the z axis, which is the transport direction. The red arrows 
mark the direction and the size of the magnetic moments within the scattering region. The small induced 
magnetic moments are overlayed by the atomic radii. The black dashed boxes illustrate the interface.
(b) The calculated spin-resolved bulk band structures of the fcc unit cell for FeVTaAl (left panel) and 
MnVTiAl (right panel). The dashed black line denotes the Fermi level that is set to zero.}
\label{fig4}
\end{figure*}

\begin{figure*}
\centering
\includegraphics[width=0.99\textwidth]{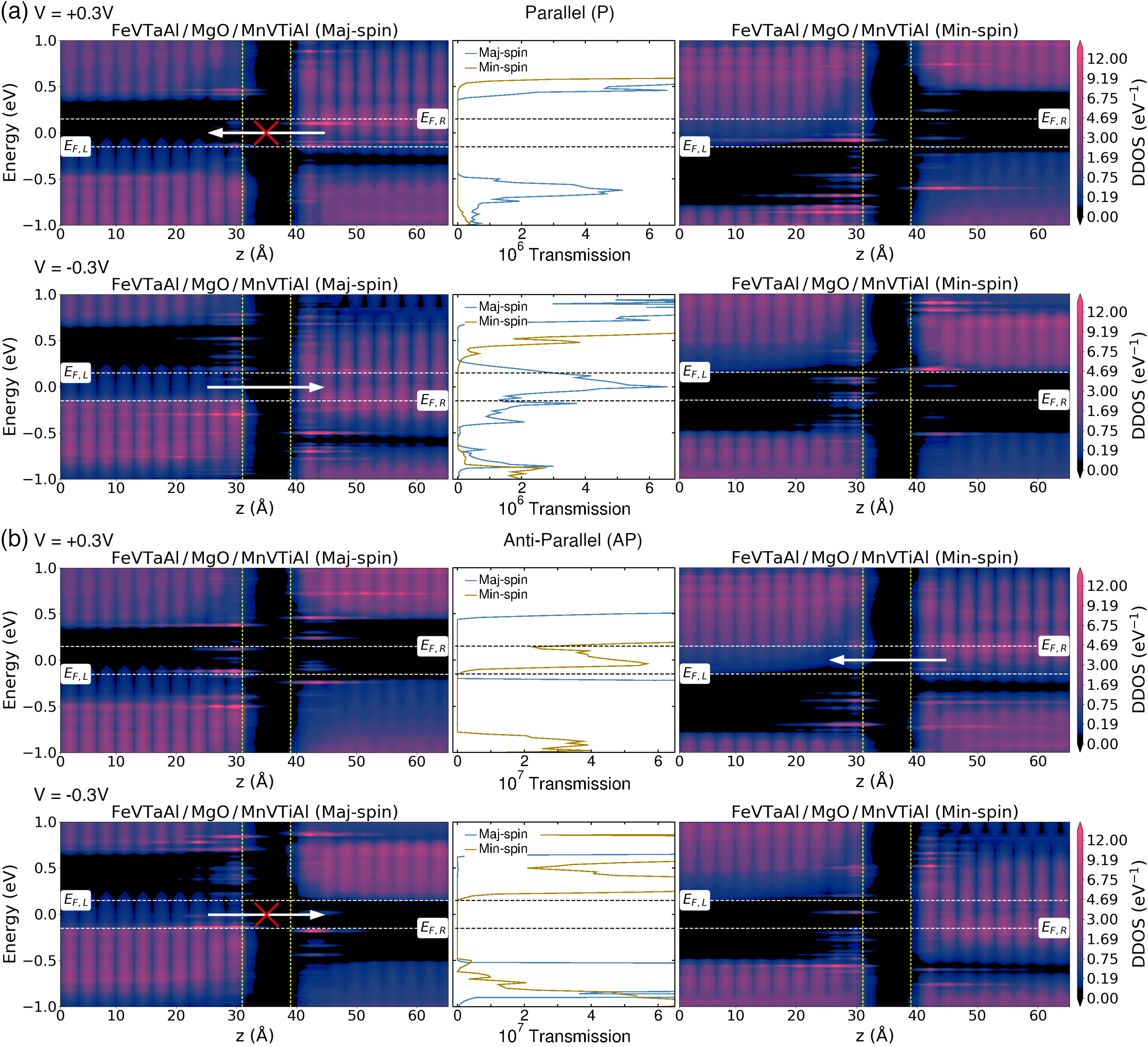}
\caption{(a) Projected device density of states (DDOS) for the majority- (left panels) and minority- (right panels) spin channel of the FeVTaAl/MgO/MnVTiAl junction for parallel orientation of the
magnetization directions of the electrodes under applied biases of +0.3 and -0.3\,V [the corresponding
atomic structure is presented in Fig.~\hyperref[fig4]{4\,(a)}]. In the middle panels we show the calculated transmission 
spectra for both spin channels. The dashed lines display the Fermi level of the left and right electrodes, 
while the vertical yellow dashed lines denote the interfaces between the electrodes and MgO. The MgO 
tunnel barrier thickness is taken to be 1.1\,nm, i.e., five monolayers. (b) The same as (a) for antiparallel
orientation of the magnetization directions of the electrodes. The majority- and minority-spin channels
are illustrated with respect to the magnetic orientation of the left electrode.}
\label{fig5}
\end{figure*}

\begin{figure*}
\centering
\includegraphics[width=0.99\textwidth]{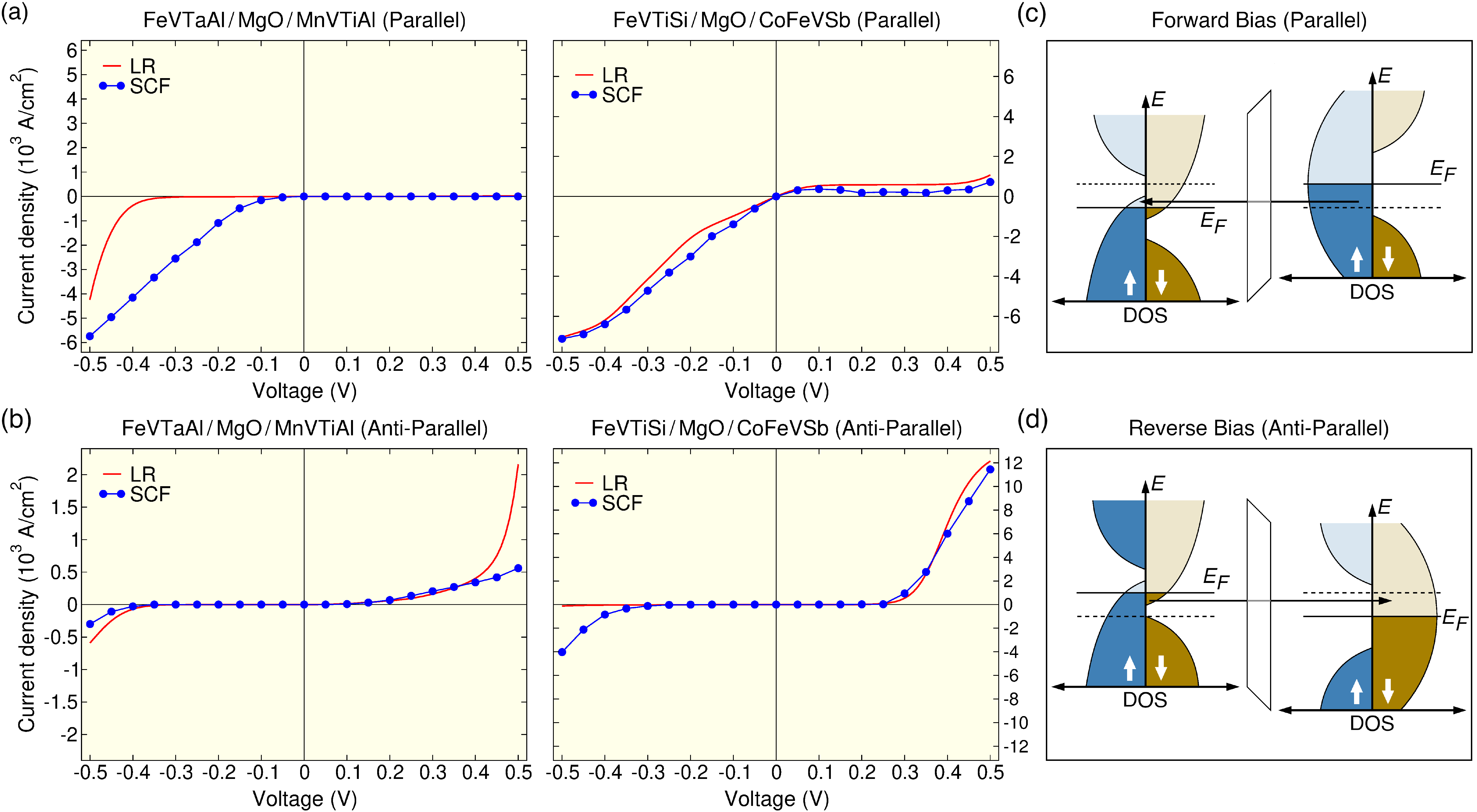}
\caption{(a) The current-voltage characteristics for the FeVTaAl/MgO/MnVTiAl (left panel) and 
FeVTiSi/MgO/CoFeVSb (right panel) junctions for five monolayers of MgO barrier thickness in the 
parallel configuration. The $I-V$ curves are calculated using both SCF and LR methods. (b) The same 
as (a) for the antiparallel alignment of the magnetization directions of the electrodes. Panels (c) and (d) 
illustrate the origin of the leakage current under forward and reverse biases for parallel and antiparallel
orientations of the magnetization direction of the electrodes, respectively.}
\label{fig6}
\end{figure*}

In \cref{sec:junction}, we qualitatively discussed the $I$-$V$ characteristics of the MTJs based on 
SGSs and HMMs using the spin dependent energy-band diagram and a simple tunnel barrier model. 
However, quantum tunneling is a very sophisticated process in real materials as it depends on 
the symmetry of the wave functions in the electrodes, their decay rate, and their matching at the 
interface. The decay rate is determined by the thickness and barrier height as well as the complex 
energy bands of the insulating material~\cite{butler2001spin,faleev2017heusler}. Therefore, fully 
\textit{ab initio} quantum transport calculations are needed to determine the $I$-$V$ characteristics of the 
MTJs based on SGSs and HMMs. We choose FeVTaAl and FeVTiSi quaternary Heusler compounds as SGS electrodes
together with MnVTiAl and CoFeVSb as HMM electrodes and construct two different types of MTJs. All four
electrode materials possess Curie temperatures above room-temperature as presented in 
\cref{tab1}. To construct the MTJs, we take the type-II SGS electrode material in the cubic 
structure and relax the tunnel barrier MgO as well as the HMM electrode material with respect to the 
in-plane lattice parameter of the first electrode. For this reason, we include the $c/a$ ratios 
for the HMM electrodes and MgO, respectively, which take the tetragonal structure in \cref{tab1}.
The atomic structure of one MTJ is illustrated in Fig.~\hyperref[fig4]{4\,(a)}. The left electrode
FeVTaAl is a SGS, the right electrode MnVTiAl is a HMM, and MgO acts as a tunnel barrier. The FeVTaAl 
(MnVTiAl) has two types of interface terminations with MgO: FeV and TaAl (MnV and TiAl). Our total 
energy calculations have shown that the FeV-MgO (MnV-MgO) termination possesses lower energy. Similarly, 
as for the second MTJ FeVTiSi/MgO/CoFeVSb (see \cref{tab1}), the FeV-MgO (CoFe-MgO) 
termination has lower energy.

For both MTJs, the thickness of the MgO tunnel barrier varies between three and six monolayers ($0.6-1.4$\,nm)
and the SGS and HMM electrodes are constructed by repeating the minimal tetragonal unit cell 5 times 
along the [001] direction. Depending on the number of MgO layers, the length of the device (screening region)
lies between 60 and 66\,{\AA}. The device is periodic in the $x-y$ plane and the $z$ direction
is the transport direction. We adjust the alignment of the magnetic moments to the $z$ axis. The 
direction and magnitude of the atomic magnetic moments of the electrode materials in the MTJ are respectively represented by 
the red arrows and their size in Fig.~\hyperref[fig4]{4\,(a)}. At both interfaces, the magnetic moments
deviate from their bulk values (see \cref{tab1}). At the FeVTaAl-MgO interface, the largest difference is 
obtained for the Fe atom whose magnetic moment increases from 0.85\,$\mu_B$ to 1.82\,$\mu_B$ while the
moment of the Ta atom changes from about 0.2\,$\mu_B$ to -0.4\,$\mu_B$. The changes at the remaining 
atoms are negligible. Similar behavior is observed for the MnVTiAl-MgO interface, where the largest deviation
occurs in the magnetic moment of the Mn atom, whose value increases from -2.42\,$\mu_B$ to -3.30$\mu_B$ 
while the magnetic moments of the other atoms remain more or less unchanged.

Next, we discuss the electronic properties of the FeVTaAl/MgO/MnVTiAl junction at zero bias, i.e., 
in equilibrium. Thus, we present the bulk band structures of both junction 
materials in Fig.~\hyperref[fig4]{4\,(b)}. The MnVTiAl compound exhibits a nearly symmetric band gap of 330\,meV 
above and 310\,meV below the Fermi level in the minority-spin channel, while FeVTaAl exhibits a type-II 
SGS behavior.
As discussed above, the strong variation of the magnetic moments at the interface implies that the
HMM and SGS properties are also lost. The loss of the HMM and SGS properties stems from two factors: 
(i) the electronic structure, i.e., the Fe and V (Mn and V) atoms at the interface possess different local atomic
environments, and thus nonbonding states can emerge close to the Fermi level; such states significantly 
reduce the spin polarization at the interface~\cite{miura2007coherent,miura2008half}; and 
(ii) charge transfer across the tunnel junction due to the work function difference of the electrodes 
(see \cref{tab1}). Since MnVTiAl exhibits the lower work function, electrons flow from the majority-spin 
(minority-spin) channel of MnVTiAl to the majority-spin (minority-spin) channel of FeVTaAl for parallel 
(antiparallel) alignment of the magnetization directions of the electrodes. When this charge redistribution 
reaches equilibrium, MnVTiAl is positively charged near the interface region, whereas FeVTaAl is negatively 
charged and, as a result, an electric dipole is induced, which affects the electronic as well as the magnetic 
properties of both electrode materials around the interface region. The loss of HMM and SGS properties at the
interface region can be seen in the device density of states (DDOS) presented in \cref{fig5} (see also Fig.~2 
in the Supplemental Material~\cite{supplement}.

The $I$-$V$ characteristics of the MTJs under consideration are calculated by using two different 
approaches: (i) the finite-bias NEGF method and (ii) a linear response approach. The latter is 
computationally much cheaper, while, however, significant differences may appear in the calculated 
$I$-$V$ curves when compared to the self-consistent NEGF calculations, as will be discussed below. 
In the middle panels of Fig.~\hyperref[fig5]{5\,(a)} and \hyperref[fig5]{5\,(b)}, we present the 
calculated transmission spectra for the FeVTaAl/MgO/MnVTiAl MTJ for applied bias voltages of $+0.3$ 
and $-0.3$\,V for the parallel and antiparallel configurations of the magnetization direction of 
the electrodes, respectively. The transmission spectrum and consequently the $I$-$V$ curves of the 
FeVTaAl/MgO/MnVTiAl MTJ displayed in \cref{fig6} can be explained on the basis of the DDOS 
[Fig.~\hyperref[fig5]{5\,(a)} and \hyperref[fig5]{5\,(b)}]. For parallel orientation of the 
magnetization directions of the electrodes, under forward bias ($V=+0.3$\,V), the transmission 
coefficient for majority-spin electrons is zero due to the gap in the type-II SGS material above 
the Fermi level. Since MnVTiAl exhibits a gap in the minority-spin channel around the Fermi energy, 
the transmission coefficient is also zero for minority-spin electrons and thus the MTJ is in the off state, 
i.e., no current flows through it under forward bias. Under reverse bias ($V=-0.3$\,V), 
majority-spin electrons of occupied states in the SGS electrode FeVTaAl can tunnel into
unoccupied states of the HMM electrode MnVTiAl through the MgO tunneling barrier and, as a 
consequence, the transmission coefficient takes a finite value. In the minority-spin channel, 
FeVTaAl possesses a gap below $E_F$ and MnVTiAl below and above the Fermi level, and hence, in 
both materials, no states are available that could contribute to a current within the applied 
voltage window. Thus, the on current of the MTJ in the parallel configuration is 100\,\% spin polarized.

Switching the magnetization direction of the electrodes from the parallel to the antiparallel configuration
results in switching the $I$-$V$ characteristics of the MTJ (see Fig.~\hyperref[fig1]{1\,(b)} and 
\hyperref[fig1]{1\,(c)}), i.e, the MTJ is in on state under forward bias, while it is in off state under 
reverse bias. In this case, the HMM electrode MnVTiAl possesses a gap in the majority-spin channel 
and thus this channel does not contribute to the current. However, in the minority-spin channel, 
electrons from the occupied states above the Fermi energy in MnVTiAl can tunnel through the MgO tunnel 
barrier into unoccupied states in FeVTaAl, and thus the transmission coefficient takes a finite value 
under forward bias, which again leads to a 100\,\% spin polarization of the current. For a
reverse bias, no current flows through the MTJ since in the majority-spin channel the HMM electrode 
MnVTiAl possesses a gap around the Fermi energy, while in the minority-spin channel the SGS
electrode FeVTaAl presents a gap below $E_F$, and hence for both spin channels the transmission 
coefficient is zero.

In Fig.~\hyperref[fig6]{6\,(a)} and \hyperref[fig6]{6\,(b)}, we present the $I$-$V$ characteristics 
of both MTJs within the finite-bias NEGF method, which will be called the self-consistent field (SCF) and 
the linear response approach (LR) for a MgO thickness of five monolayers (1.1\,nm). As seen for the parallel 
orientation of the magnetization direction of the electrodes, both MTJs are in the off state under forward 
bias and in the on state under reverse bias. However, this might be seen as contradicting to the conventional 
$p-n$ diodes, in which the diode is in the on state under forward bias. In our case this is a matter of the 
construction of the MTJ, i.e., by exchanging the electrode materials, one obtains the $I$-$V$ 
characteristics of conventional diodes. In the SCF calculations, we obtain a monotonic increase of the 
current $I$ with bias voltage $V$ for both MTJs with zero turn-on voltages $V_T$ for both FeVTaAl/MgO/MnVTiAl
and FeVTiSi/MgO/CoFeVSb junctions in the parallel configuration, respectively. Switching the magnetization 
direction of the HMM electrode from parallel to antiparallel results in switching the $I$-$V$ 
characteristics of the MTJs as shown in Fig.~\hyperref[fig6]{6\,(b)}. Both MTJs are now in the on state under 
forward bias, while they are in the off state under reverse bias. In contrast to the parallel alignment of 
the magnetization directions, in this case, the turn-on voltage $V_T$ for the FeVTiSi/MgO/CoFeVSb junction is 
large, i.e., $0.25$\,V, which can be understood on the basis of the DDOS presented in the Supplemental 
Material~\cite{supplement}. The large work function difference of the electrode materials (FeVTiSi and 
CoFeVSb) gives rise to a band bending in the energy-band diagram of this MTJ and as a consequence one 
obtains an effectively thick tunnel barrier for small bias voltages, which leads to a large turn-on 
voltage under forward bias. Furthermore, the on state currents for parallel and antiparallel configurations
of the same MTJ is also quite different. For instance, in the FeVTaAl/MgO/MnVTiAl junction for the parallel 
configuration the on state current is one order of magnitude larger than the corresponding current in the
antiparallel configuration, while in the FeVTiSi/MgO/CoFeVSb junction the situation is different; here,
the on state current is a factor of 2 smaller in the parallel configuration. For comparison, the 
$I$-$V$ curves obtained from the linear-response approach have been 
included in Fig.~\hyperref[fig6]{6\,(a)} and \hyperref[fig6]{6\,(b)}. As seen, qualitatively the linear 
response current follows the SCF results with some differences such as the turn-on voltage in the case 
of the FeVTaAl/Mgo/MnVTiAl junction in the parallel configuration and the overestimated leakage current in 
the case of the FeVTiSi/MgO/CoFeVSb junction also for the parallel configuration. We do not expect a 
quantitative agreement between these approaches because for the linear response method one assumes a 
bias-independent transmission spectrum and thus this method is not capable of an accurate description 
of the $I$-$V$ characteristics. The zero-bias transmission spectrum for the linear response calculations and a discussion of the charge accumulation at the interfaces as well as a detailed analysis of the symmetry character of the band structure of the electrode materials 
of both MTJs can be found in the Supplemental Material~\cite{supplement}.

We would now like to comment on the off state leakage currents of both MTJs. In principle, at zero
temperature, one would obtain a zero off state current for a perfect SGS electrode. However, in 
our MTJs both SGS electrodes, FeVTaAl and FeVTiSi, possess a sizeable band overlap between the 
valence and conduction bands of opposite spin channels around $E_F$ as schematically illustrated 
in Figs.~\hyperref[fig6]{6\,(c)} and \hyperref[fig6]{6\,(d)} (see also the supplemental material of 
Ref.~\onlinecite{aull2019ab} for the DOS). For parallel (antiparallel) aligned magnetization 
directions of the electrodes, band overlaps allow majority-spin (minority-spin) electrons to tunnel 
from the occupied states of the HMM (type-II SGS) electrode through the insulating region into 
unoccupied states of the type-II SGS (HMM) material. Since FeVTiSi possesses an overlap of 150\,meV 
whereas the overlap in FeVTaAl amounts to just 60\,meV, a larger leakage current arises in the 
FeVTiSi/MgO/CoFeVSb junction. At zero temperature, the obtained on:off current ratios of both MTJs 
at $\pm 0.3\,V$ vary between $10^2$ and $10^7$.

\begin{figure*}[!t]
\centering
\includegraphics[width=0.98\textwidth]{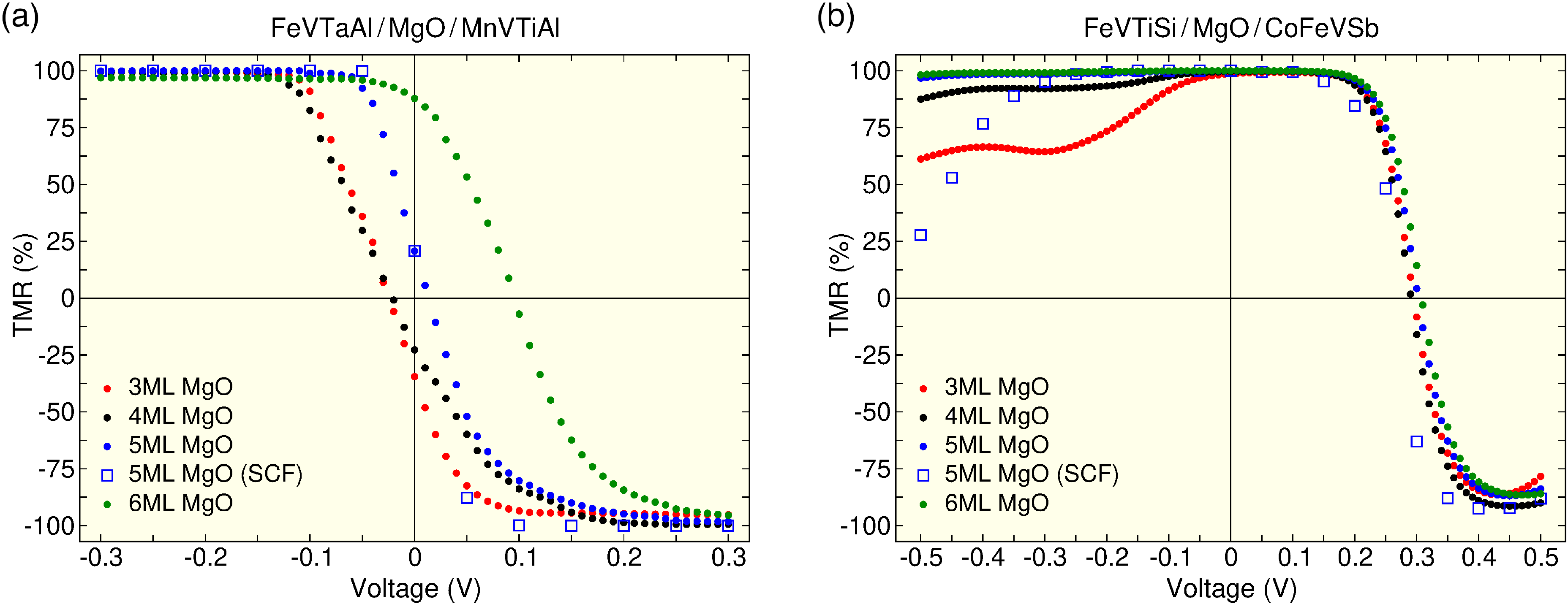}
\caption{(a) Voltage dependence of the TMR ratio of the FeVTaAl/MgO/MnVTiAl MTJ for different
MgO thicknesses calculated within the linear response approach. For the case of five monolayers (MLs)
of MgO thickness, the results are compared with SCF calculations. (b) The same as (a) for the
FeVTiSi/MgO/CoFeVSb MTJ.}
\label{fig7}
\end{figure*}

In contrast to conventional semiconductor diodes ($p-n$ diode, Schottky diode, Zener diode), in which the 
rectification bias voltage window (or reverse bias breakdown voltage of the diode) varies between $3-200$\,V,
in the present MTJs, this voltage window is limited by the spin gap of the HMM and SGS Heusler compounds. 
In analogy to conventional semiconductor diodes, we can express the breakdown voltage for the parallel and
antiparallel configurations as $V_B^{\textrm{P}}=\min \left\{(\textrm{SG}_{\textrm{SGS}},E_F^{\textrm{A}}),
(\textrm{SG}_{\textrm{HMM}},E_F^{\textrm{B}})\right\}$ and $V_B^{\textrm{AP}}=\min \left\{(\textrm{SG}_{\textrm{SGS}},E_F^{\textrm{B}}),(\textrm{SG}_{\textrm{HMM}},E_F^{\textrm{A}})\right\}$, 
where $(\textrm{SG}_{\textrm{SGS}},E_F^{\textrm{A}})$ and $(\textrm{SG}_{\textrm{HMM}},E_F^{\textrm{B}})$
stand for the spin gaps of the SGS and HMM electrodes above and below the Fermi energy, respectively. 
Using the spin gap values of the SGSs and HMMs from Ref.~\onlinecite{aull2019ab}, one gets breakdown voltages $V_B^{\textrm{P}}$ 
($V_B^{\textrm{AP}}$) of $0.31$ and $-0.34$\,V ($0.33$ and $-0.30$\,V) for the parallel (antiparallel) configurations 
of the FeVTaAl/MgO/MnVTiAl and FeVTiSi/MgO/CoFeVSb MTJs, respectively. Since the estimation of the breakdown 
voltage is based on the DOS picture of the materials, the calculated $V_B^{\textrm{P}}$ ($V_B^{\textrm{AP}}$) values
can differ substantially since, as mentioned before, in tunneling processes the bands along the transport direction,
their symmetry character, and their matching across the interface play a decisive role. Indeed, the actual calculated 
$V_B^{\textrm{P}}$ values in \cref{fig6} are larger than the simple estimated ones for the parallel configuration, while
for the antiparallel configuration, the calculated $V_B^{\textrm{AP}}$ values are more close to the estimated ones. 
However, the simple estimated values set the lower limit of the breakdown voltages $V_B^{\textrm{P}}$ and $V_B^{\textrm{AP}}$.

Recently Maji and Nath reported the fabrication of a MTJ based on Heusler compounds as electrode materials. The junction is composed of HMM Co$_2$MnSi, SGS Ti$_2$CoSi, and a 3-nm 
MgO tunnel barrier~\cite{maji2022demonstration}. The authors demonstrated a reconfigurable diode effect with 
a high on:off ratio and a high TMR ratio that decreases with 
increasing temperature. Moreover, the breakdown voltage of the MTJ under reverse bias was reported to be
around -0.5\,V, which is basically the spin gap of the Co$_2$MnSi compound. Indeed, this is the first experimental 
demonstration of  
the concept that we proposed in 2019~\cite{sasioglu2019proposal}. Note that Maji and Nath~\cite{maji2022demonstration} used Ti$_2$CoSi as a SGS electrode; however, this material exhibits 
type-III SGS behavior in a simple DOS picture~\cite{skaftouros2013search}, whereas in tunneling experiments it 
behaves like a type-II SGS due to the reasons we discussed above. A detailed discussion of the experiments 
of Maji and Nath is beyond the scope of the present paper since we were aware of this work after the completion 
of the present paper. However, we are planning to consider the MTJ of Ref.~\onlinecite{maji2022demonstration} in 
a separate future study.

The $I$-$V$ characteristics discussed above as well as the TMR ratio, which we discuss below,
in our MTJs are calculated for zero temperature. The temperature effects (non-spin-flip thermal 
excitations; see \cref{fig3}) are usually included in NEGF transport calculations of semiconductor 
devices via the Fermi-Dirac distribution function. However, due to the technical limitation of 
the \textsc{QuantumATK} package, as discussed in detail in Ref.~\onlinecite{sasioglu2020half} for 
spintronic materials, we neglect these thermal excitations in transport calculations. Moreover,
besides high-energy non-spin-flip thermal excitations, temperature affects the magnetic and 
electronic structure of the SGSs and HMMs via Stoner excitations and magnons or collective spin 
waves. In type-II SGSs, electrons around the Fermi energy can be excited via spin flip with 
a nearly vanishing amount of energy [see Fig.~\hyperref[fig2]{2\,(a)}]. Such excitations 
are known as single-particle Stoner excitations and occupy, in our case, the unoccupied 
minority-spin states above $E_F$. As a consequence, these electrons contribute to a leakage 
current in the antiparallel orientation of the magnetization direction of the type-II SGS 
and HMM electrode. On the other hand, due to the existence of a gap in HMMs, Stoner excitations
are not allowed in these materials. Nevertheless, at finite temperatures, electron-magnon 
interactions might give rise to the appearance of nonquasiparticle states in the spin gap above 
the Fermi level of HMMs~\cite{katsnelson2008half}. As a consequence, these states reduce the spin polarization 
of the HMM material and thus influence its transport properties. Furthermore, defects 
at the interface might also affect the characteristics of the SGSs and HMMs and contribute to a 
leakage current and reduce the on:off ratio and TMR effect.

Finally, we discuss the TMR effect in the MTJs under study. As mentioned above, the 
reconfigurable diode effect gives rise to an inverse TMR effect in this type of MTJ. The voltage 
dependence of the TMR ratio for both MTJs is presented in Fig.~\hyperref[fig7]{7\,(a)} and \hyperref[fig7]{7\,(b)}
for four different MgO tunnel barrier thicknesses. Because of the computational efficiency, we stick 
here to the linear response approach; however, for five monolayers of MgO tunnel barrier thickness,
we compare the obtained results with the SCF method. For negative bias voltages (reverse bias), both 
MTJs present a positive TMR effect, while at a certain applied bias voltage, due to the unique band structure
of the SGS electrode, the TMR changes its sign to a negative value, and thus the MTJs exhibits an 
inverse TMR effect. In principle, for a perfect SGS electrode material, one expects a sharp transition 
from positive to negative TMR values at zero bias voltage as displayed in Fig.~\hyperref[fig1]{1\,(a)}; 
however, in the present MTJs, this transition takes place in a finite voltage window and the transition 
point is shifted to finite voltages especially in the FeVTiSi/MgO/CoFeVSb tunnel junction. Two parameters 
are mainly responsible for the behavior of the TMR curves. These are the on:off current ratio, which 
reduces the TMR ratio, and the threshold voltage $V_T$, which causes a voltage shift of the transition 
point. Like in $I$-$V$ curves, the spin gap of the electrode materials plays an essential role for the
TMR ratio and its sign. For instance, in the  FeVTiSi/MgO/CoFeVSb tunnel junction the high TMR is obtained in 
a very small voltage window, especially for negative voltages and the TMR ratio is significantly reduced 
for voltages beyond $-0.3$\,V, which is more or less the spin gap of the HMM CoFeVSb material.

\section{\label{sec:concl}SUMMARY AND OUTLOOK}

MTJs based on Fe, Co, and CoFeB, as well as HMM Heusler compounds, have been extensively studied in spintronics 
for magnetic memory and magnetic logic applications. Despite their high TMR ratios, especially the MTJs 
based on HMMs, conventional MTJs do not exhibit current rectification, i.e., a diode effect. A  
MTJ concept has been proposed in Ref.~\onlinecite{sasioglu2019proposal}, which exhibits reconfigurable current 
rectification together with an inverse TMR effect. This MTJ concept was based on HMMs and SGSs and 
it has recently been demonstrated experimentally using Heusler compounds~\cite{maji2022demonstration}. In 
the present work, by employing the state-of-the-art DFT and NEGF methods, we design two different MTJs 
based on the type-II SGS and HMM quaternary Heusler compounds FeVTaAl, FeVTiSi, MnVTiAl, and CoVTiSb. We 
show that both MTJs [FeVTaAl(001)/MgO/MnVTiAl(001) and FeVTiSi(001)/MgO/CoFeVSb(001)] exhibit a 
current rectification with a relatively high on:off ratio of up to 10$^7$. We show that in contrast 
to conventional semiconductor diodes, such as the $p-n$ junction diode or Schottky diode, the rectification bias 
voltage window (or breakdown voltage) of these MTJs is limited by the spin gap of the HMM and SGS Heusler 
electrode material in agreement with recent experiments. A unique feature of the present MTJs is that 
they can be configured dynamically, i.e., depending on the relative orientation of the magnetization 
direction of the electrodes, the MTJ allows electrical current to pass either in one or the other direction. 
This feature gives rise to an inverse TMR effect in such devices. The inverse TMR effect has been investigated 
as a function of the MgO tunnel barrier thickness. We find that the sign change of the TMR from a positive 
to a negative value takes place not at zero bias voltage, but small finite voltages, which can be explained 
by the on:off ratio (leakage current) and threshold voltage $V_T$ of the MTJs. Moreover, like in $I$-$V$ 
curves, the spin gap of the electrode materials plays an essential role in the TMR ratio and its sign.

The current nonvolatile magnetic memory technology (STT MRAM and beyond) and several magnetic logic 
proposals utilize conventional MTJs that have limited functionality. The 
MTJs based on HMMs and
SGSs studied in the present paper provide major advantages over conventional MTJs and open  
ways to magnetic memory and logic concepts. For instance, these MTJs might be of particular interest for 
high-density 3D cross-point STT MRAM applications as they eliminate the need for an additional 
selection device such as a MOSFET transistor or a $p-n$ diode. Apart from memory applications, the MTJs 
constitute the basic building blocks of the three-terminal magnetic tunnel transistors with unique 
properties as discussed in Ref.~\onlinecite{sasioglu2019proposal}. Moreover, the present MTJs also open 
the way to logic-in-memory computing, i.e., storing and processing the data within the same chip and 
thus providing an opportunity to explore  
computing architectures beyond the classical von Neumann architecture.

\acknowledgements

This work is supported by SFB CRC/TRR 227 of Deutsche Forschungsgemeinschaft (DFG) 
and by the European Union (EFRE) under Grant No. ZS/2016/06/79307.

\bibliography{bibliography}

\end{document}


\title{Supplemental Material}

\author{T. Aull}
\author{E. \c{S}a\c{s}{\i}o\u{g}lu}
\author{N. F. Hinsche}
\author{I. Mertig}

\affiliation{Institute of Physics, Martin Luther University Halle-Wittenberg, D-06120 Halle (Saale), Germany}

\maketitle

\thispagestyle{fancy}

\begin{figure}[b]
    \centering
    \includegraphics[width=0.62\textwidth]{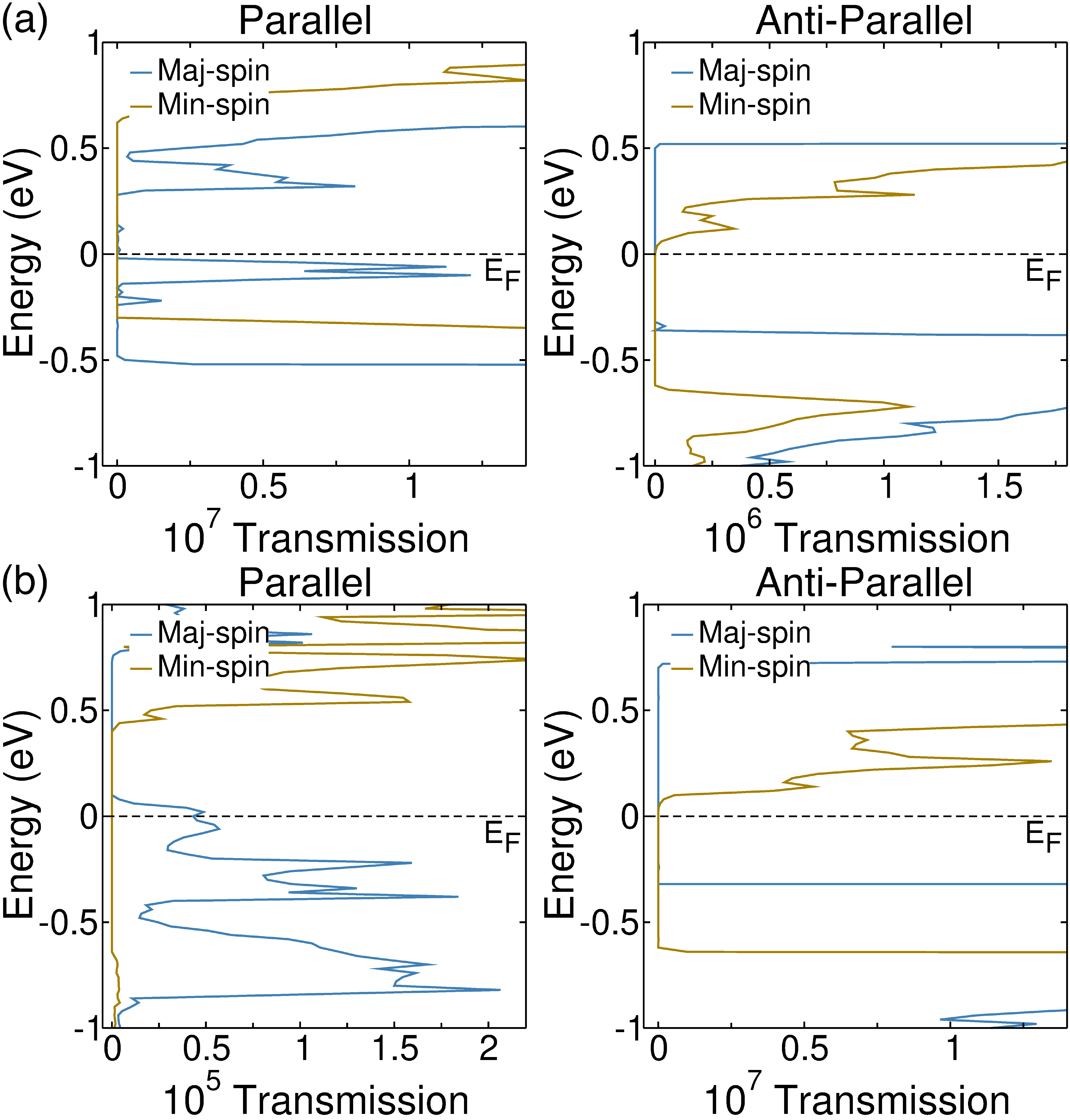}
    \vspace{-5mm}
    \caption{Zero-bias transmission spectrum for majority spin and minority spin electrons in parallel (left panel) and anti-parallel (right panel) alignment of the magnetization direction of the (a) FeVTaAl/MgO/MnVTiAl and (b) FeVTiSi/MgO/CoFeVSb junction for 5 monolayers of MgO as tunnel barrier. The black dashed line displays the Fermi level which is set to 0.}
    \label{fig1}
\end{figure}

\begin{figure}[!b]
    \centering
    \includegraphics[width=0.99\textwidth]{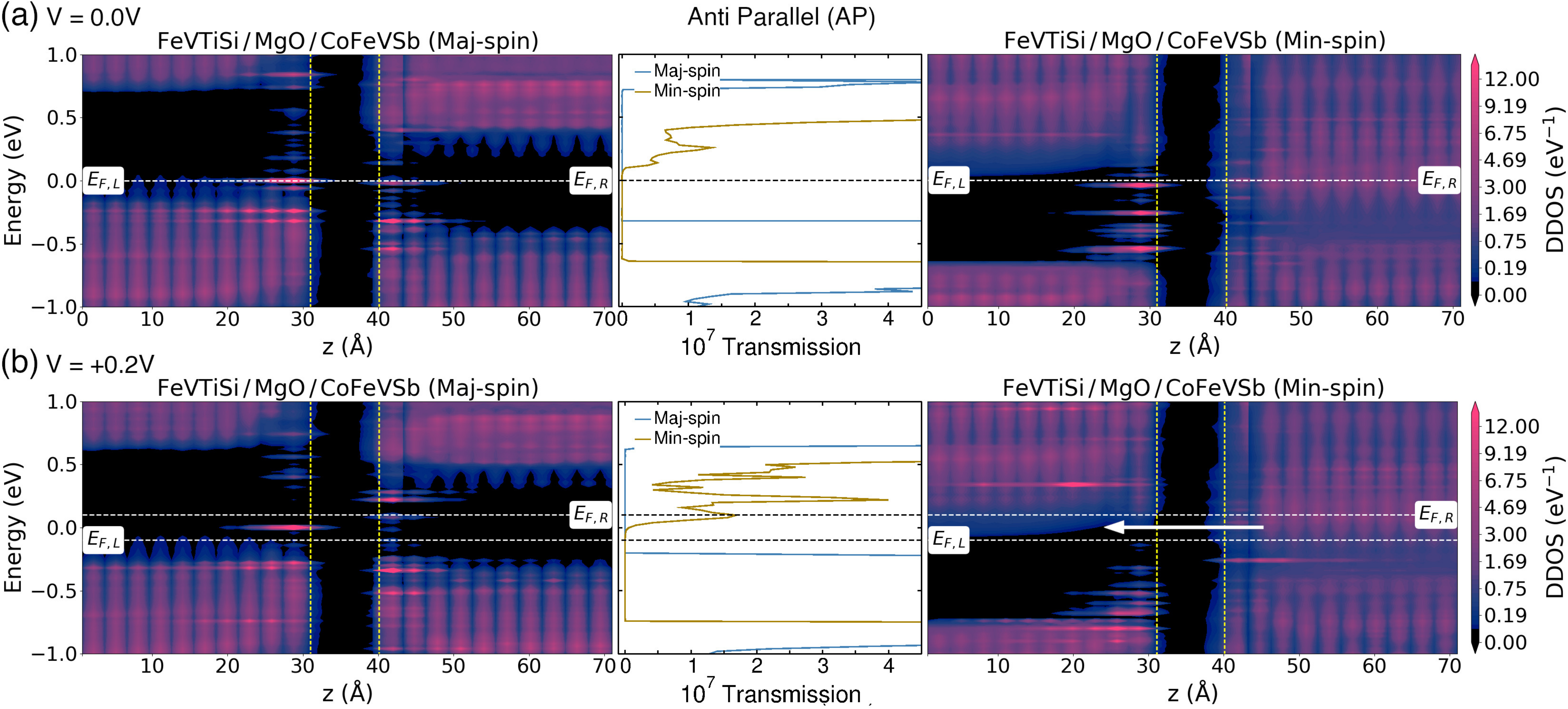}
    \vspace{-5mm}
    \caption{Projected device density of states (DDOS) for the majority (left panel) and minority (right panel) spin channel of the FeVTiSi/MgO/CoFeVSb junction for anti-parallel orientation of the magnetization directions of the electrodes (a) at equilibrium (zero bias) and (b) under an applied bias of voltage of $+0.2$\,V. In the middle panels we present the transmission spectrum for both spin channels. The dashed lines displays the Fermi level of the left and right electrode while the vertical yellow dashed lines denote the interfaces between the electrodes and MgO. The MgO tunnel barrier thickness is taken to be 1.1\,nm, i.e., five monolayers.}
    \label{fig2}
\end{figure}

In the main text, we discuss the current-voltage characteristics and tunnel magnetoresistance effect for the 
magnetic tunnel junctions (MTJs) based on quaternary Heusler compounds FeVTaAl, FeVTiSi, MnVTiAl, and CoVTiSb.   
In this supplementary part, we provide the zero-bias transmission spectrum for both MTJs (FeVTaAl/MgO/MnVTiAl (Fig.~\hyperref[fig1]{1\,(a)})  
and FeVTiSi/MgO/CoFeVSb (Fig.~\hyperref[fig1]{1\,(b)})) and give a table listing the spin gap values below and above the Fermi level for the spin-gapless and half-metallic Heusler compounds (\cref{tab1}).
In addition, we show the device density of states (DDOS), as well as the transmission spectrum of the FeVTiSi/MgO/CoFeVSb 
junction in anti-parallel configuration at equilibrium (zero bias) (Fig.~\hyperref[fig2]{2\,(a)}) and for an applied bias voltage of $+0.2$\,V (Fig.~\hyperref[fig2]{2\,(b)}), and provide a detailed symmetry analysis of the band character of both magnetic tunnel junctions.

Finally, we present and discuss the charge accumulation across the FeVTaAl/MgO/MnVTiAl junctuion under zero bias and an applied bias voltage of $\pm 0.3$\,V.

\begin{table}[!t]
    \centering
    \caption{Spin gap values of spin-gapless semiconductors and half-metallic magnets. The values are taken from Ref.~61 in the main text.}
    \begin{ruledtabular}
    \begin{tabular}{lcc}
         Material & Spin gap below $E_F$ [meV] & Spin gap above $E_F$ [meV]\\
         \hline
         MnVTiAl & 314 & 331 \\
         CoFeVSb & 340 & 206 \\
         FeVTaAl & 636 & 466 \\
         FeVTiSi & 702 & 691 \\
    \end{tabular}
    \end{ruledtabular}
    \label{tab1}
\end{table}

To get more insight into electronic transport calculations within the ballistic limit, the $k$-resolved transmission coefficients can be approximated as 
\begin{equation*}
T^{\sigma, \nu}(k_{\|}) \approx t^{\sigma, \nu}_L(k_{\|}) \cdot \exp{[- 2 \kappa(k_{\|}) d]} \cdot t^{\sigma, \nu}_R(k_{\|}).
\end{equation*}

Here $t_L$ and $t_R$ are the interface transmission functions of the left and right interface, representing the spin $\sigma$, the symmetry character $\nu$ of the wavefunction \cite{Wigner:1936}, and the amount of carriers available at $k_\|$, i.e., the density of states $n$ \cite{Belashchenko:2004}. The exponential factor accounts for the decay of the electronic wave function within the barrier of thickness $d$ with a decay rate $\kappa(k_{\|})$. The latter is given by the imaginary part of the wave vector defined by the complex band structure of the barrier material \cite{Hinsche:2010}. 

Various scenarios can occur. In the optimal case, left and right leads offer many states ($n \gg 1$) close to $E_F$ with identical or compatible symmetry characters and match with barrier states, again with compatible symmetry. In the best case, the latter states have very small decay rates ($\kappa \rightarrow 0$). If all of these arguments hold, the current flow through the tunnel junction will be large, almost metallic. 
On the other hand, if the symmetry characters of states on the left and right lead do not match, the current will vanish, even though the decay rate might be small. If the symmetry of the states match, but the decay rates are rather large, a small current will pass the device; presumably larger than in the before mentioned non-matching case, but noticeably smaller than in the first optimal case.

\begin{figure}[htb!]
 \includegraphics[width=0.95\textwidth]{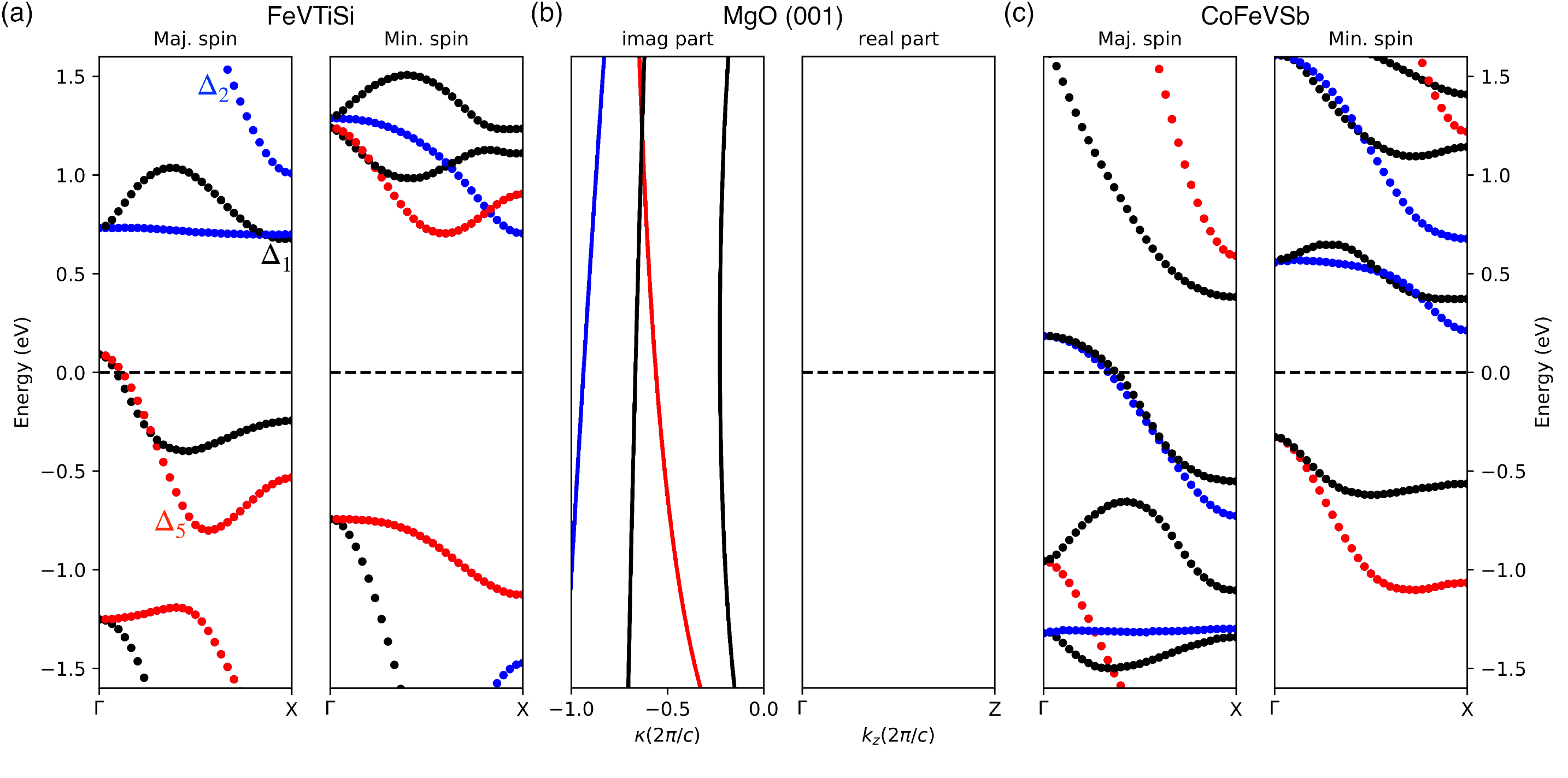}
 \caption{Real band structure of the ferromagnetic leads FeVTiSi (a) and CoFeVSb (c) along the $\Gamma-X$ high symmetry line. (b) depicts the complex band structure of the MgO tunneling barrier. Shown are the real part along the $\Gamma-X/Y/Z$ high symmetry line, as well as the smallest imaginary parts of the complex bands at $\Gamma$. The color code and the symbols relate to the crystallographic notation for the band symmetry representations. The fcc bulk systems are discussed for all cases.
 The black dashed lines denote the Fermi level.}
 \label{fig3}
\end{figure}

We will use these approximations to qualitatively affirm the results of our \textit{ab initio} calculations in the main manuscript. 

As shown in Fig.~\hyperref[fig3]{3\,(b)} MgO, offers three mentionable evanescent states within the fundamental band gap. In general, they differ in symmetry character and decay length. Close to the $\Gamma$-point, it is $\kappa_{\Delta_1} < \kappa_{\Delta_5} < \kappa_{\Delta_2}$. So, lead states matching symmetry character $\Delta_1$ would be the most desirable as these states would decay less and allow for a large current. 

We start the discussion for the junctions in parallel configuration. In the case of the FeVTiSi/MgO/CoFeVSb junction (Fig.~\ref{fig3}), we recognize states close to $E_F$ only for the majority carriers. Around the $\Gamma$-point states of symmetry characters $\Delta_1, \Delta_5$ for FeVTiSi, and $\Delta_1, \Delta_2$ for CoFeVSb can be found. 
Obviously, $t^{\uparrow, \Delta_5}_L(\Gamma) \times t^{\uparrow, \Delta_2}_L(\Gamma) \approx 0$, and thus only the $\Delta_1$ states match in symmetry, which in turn are also the states with the smallest decay rate within MgO. Hence, a strong spin-polarized current, dominated by $\Delta_1$-states around $\Gamma$, will be driven across the junction.

\begin{figure}[b!]
 \includegraphics[width=0.95\textwidth]{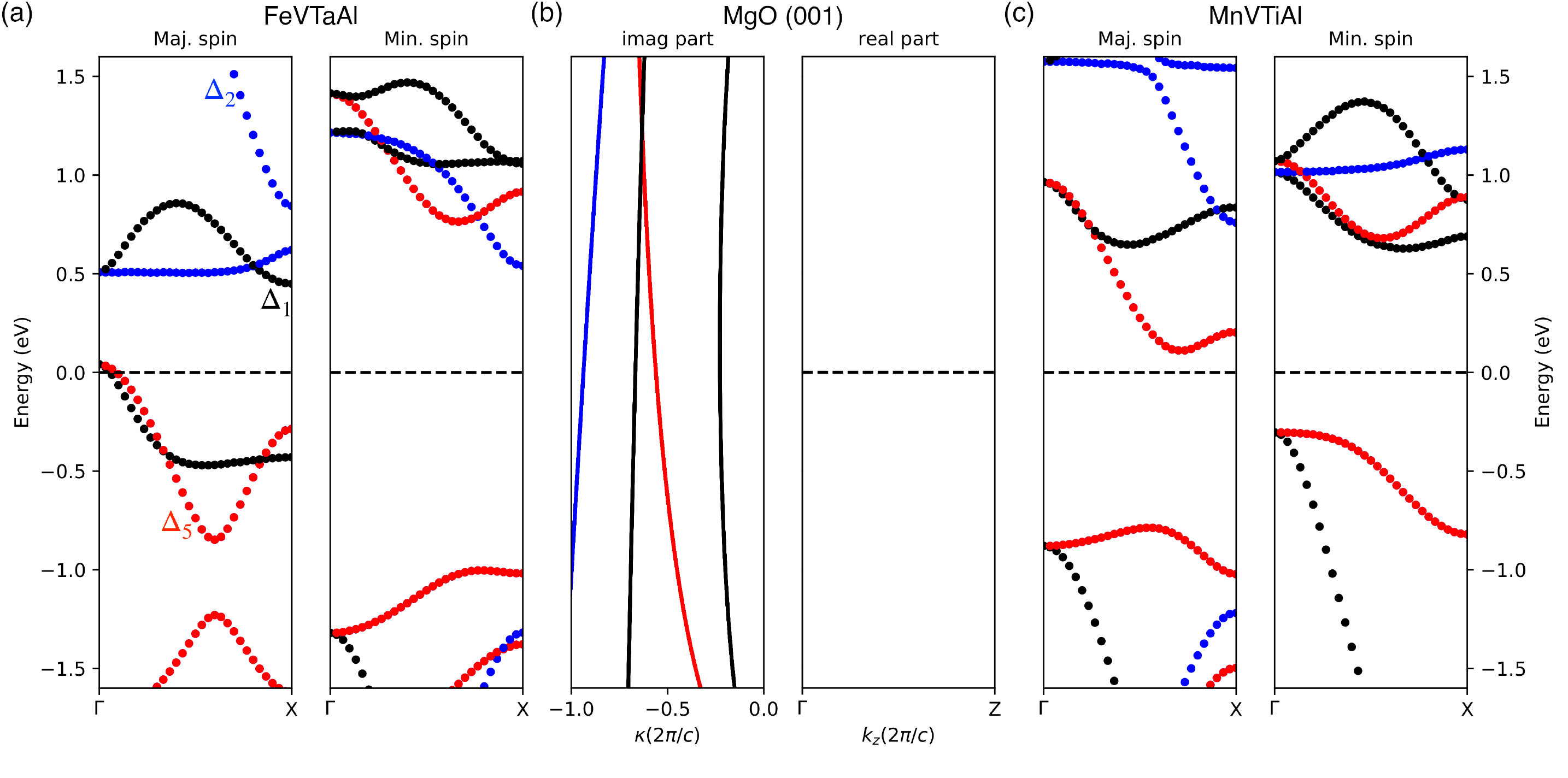}
 \caption{Real band structure of the ferromagnetic leads FeVTaAl (a) and MnTiVAl (c) along the $\Gamma-X$ high symmetry line. (b) depicts the complex band structure of the MgO tunneling barrier. Shown are the real part along the $\Gamma-X/Y/Z$ high symmetry line, as well as the smallest imaginary parts of the complex bands at $\Gamma$. The color code and the symbols relate to the crystallographic notation for the band symmetry representations. The fcc bulk systems are discussed for all cases. 
 The dashed black lines mark the Fermi level.}
\label{fig4}
\end{figure}

\begin{figure}[t!]
\includegraphics[width=0.95\textwidth]{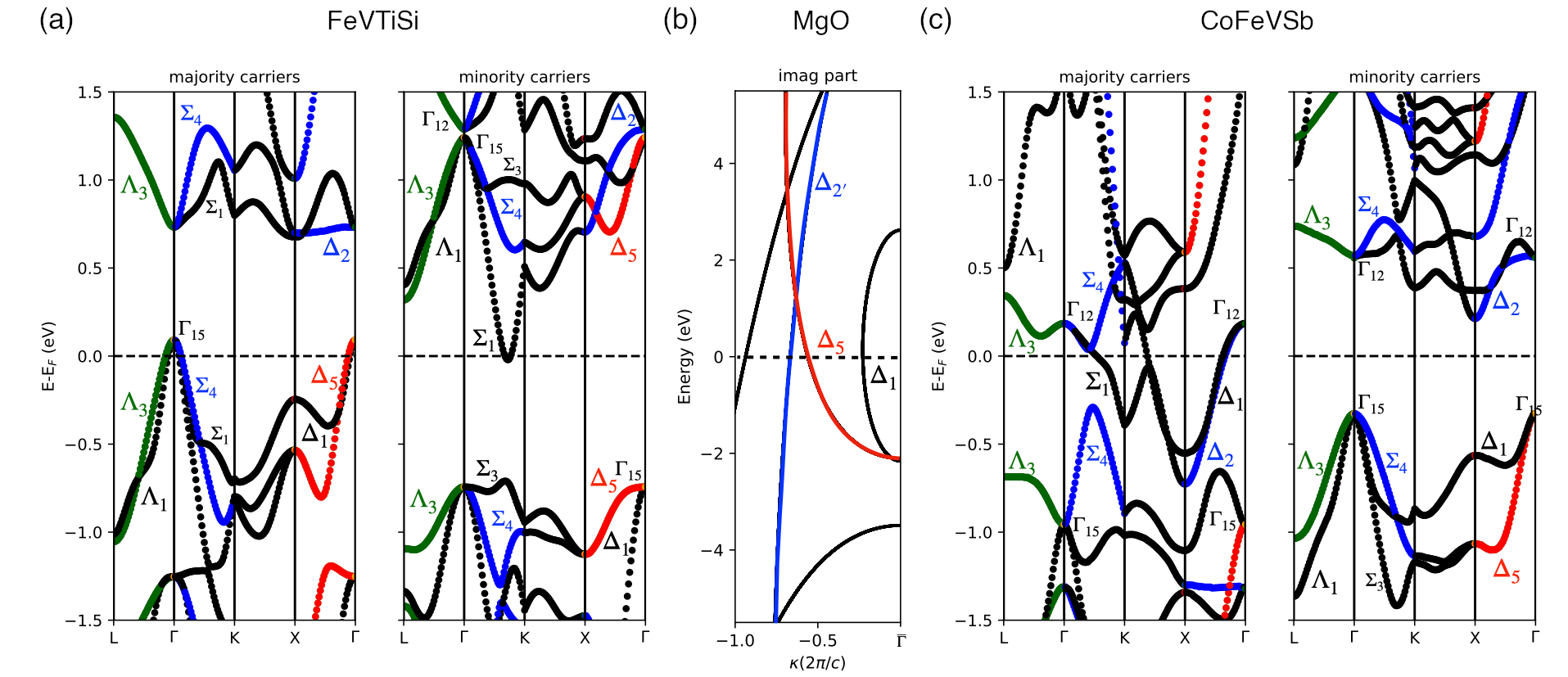}
 \caption{Real band structure of the ferromagnetic leads FeVTiSi (a) and CoFeVSb (c) along high symmetry lines. (b)depicts the smallest imaginary parts of the complex bands at $\Gamma$ for MgO. The color code and the symbols relate to the crystallographic notation for the band symmetry representations. The fcc bulk systems are discussed for all cases. Note that the energy axis slightly differ for MgO and both leads. Dashed lines mark the Fermi level.}
 \label{fig5}
\end{figure}

\begin{figure}[!b]
\includegraphics[width=0.95\textwidth]{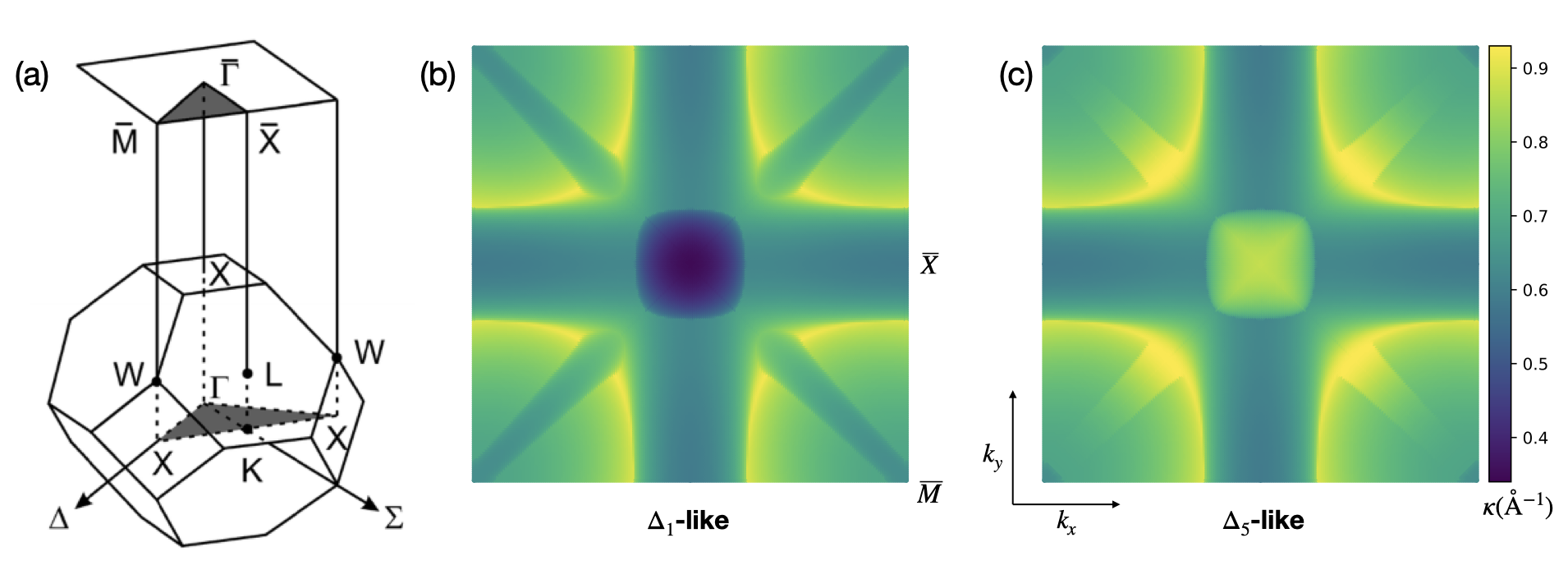}
 \caption{(a) Bulk fcc Brillouin zone and projected (001) surface/interface Brillouin zone. (b) and (c) show the two smallest imaginary parts of the complex bands in the 001 interface Brillouin zone for MgO. The Brillouin zone spans from $[-\pi/a, \pi/a]$.}
 \label{fig6}
\end{figure}

\begin{figure}[t!]
\includegraphics[width=0.95\textwidth]{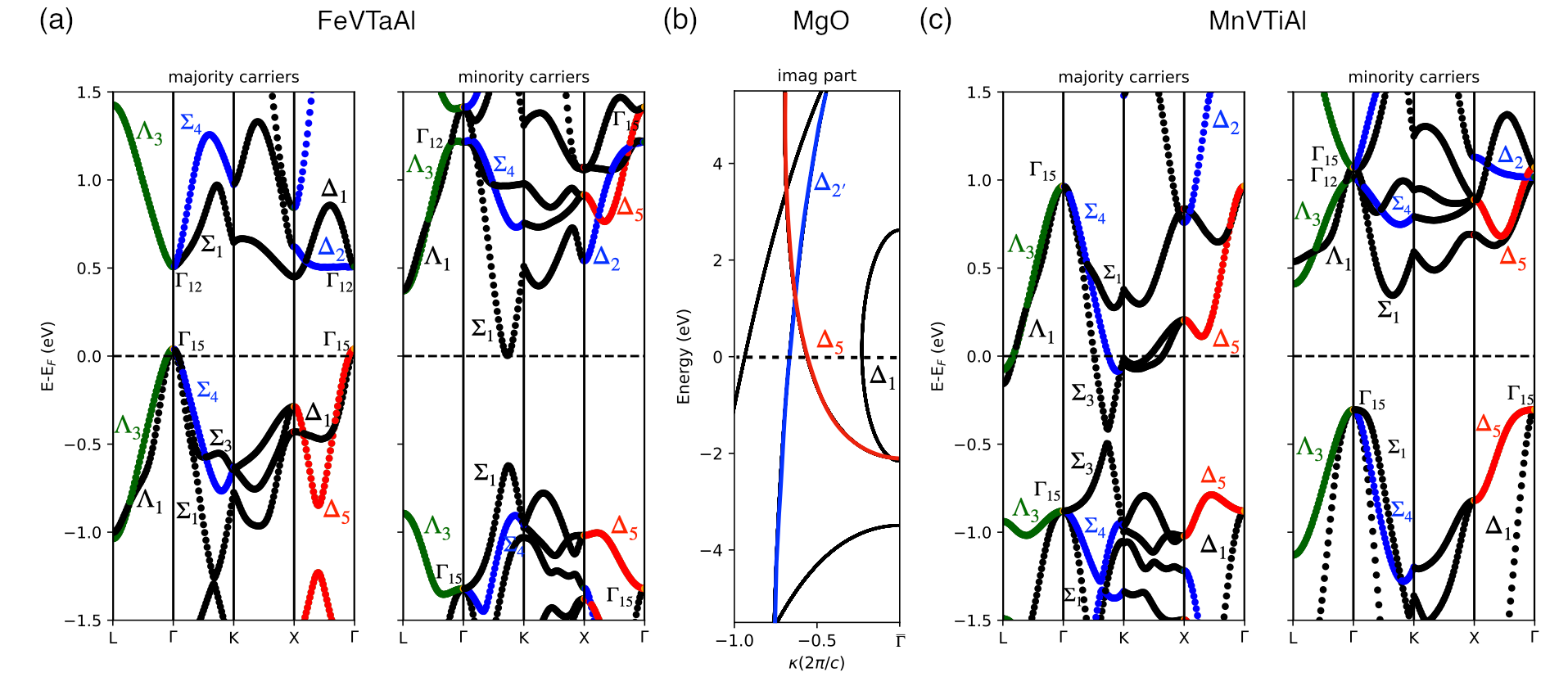}
 \caption{Real band structure of the ferromagnetic leads FeVTaAl (a) and MnTiVAl (c) along high symmetry lines. (b) depicts the smallest imaginary parts of the complex bands at $\Gamma$ for MgO. The color code and the symbols relate to the crystallographic notation for the band symmetry representations. The fcc bulk systems are discussed for all cases. Note that the energy axis slightly differ for MgO and both leads. The dashed black lines mark the Fermi level.}
 \label{fig7}
\end{figure}

\begin{figure}[!b]
    \centering
    \includegraphics[width=0.95\textwidth]{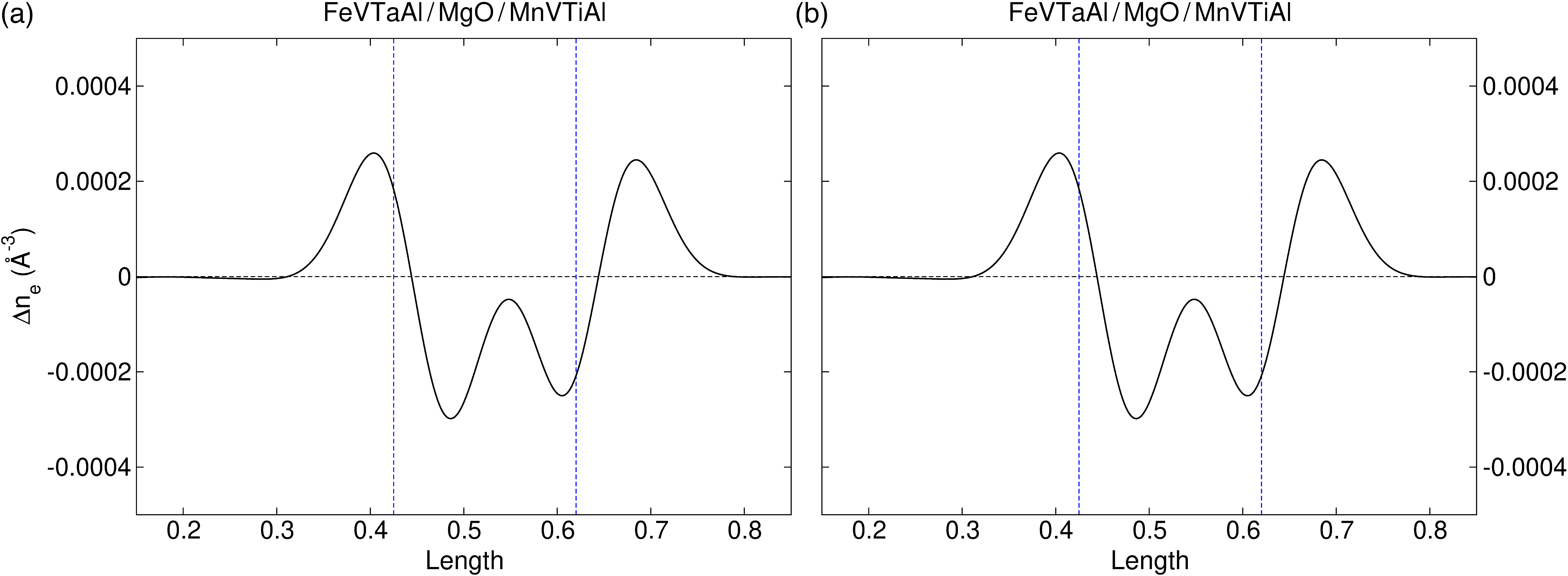}
    \caption{Electron difference density as function of the length of the FeVTaAl/MgO/MnVTiAl junction in (a) parallel and (b) anti-parallel orientation of the magnetization direction of the electrodes for 5ML of MgO at zero bias. The dashed blue lines denote the interfaces with MgO and the dashed black line the zero line.}
    \label{fig8}
\end{figure}

\begin{figure}[t!]
    \centering
    \includegraphics[width=0.95\textwidth]{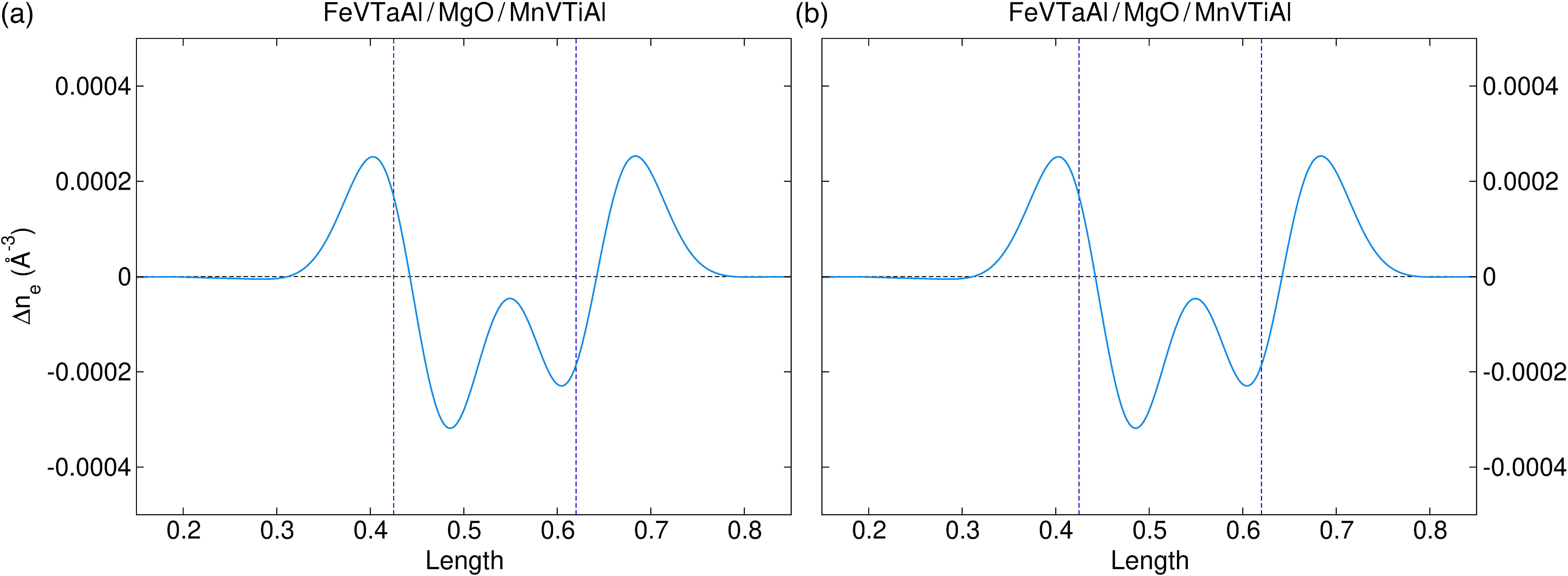}
    \caption{Electron difference density as function of the length of the FeVTaAl/MgO/MnVTiAl junction in (a) parallel and (b) anti-parallel orientation of the magnetization direction of the electrodes for 5ML of MgO at an applied bias of $0.3$\,V. The dashed blue lines denote the interfaces with MgO and the dashed black line the zero line.}
    \label{fig9}
\end{figure}

\begin{figure}[b!]
    \centering
    \includegraphics[width=0.95\textwidth]{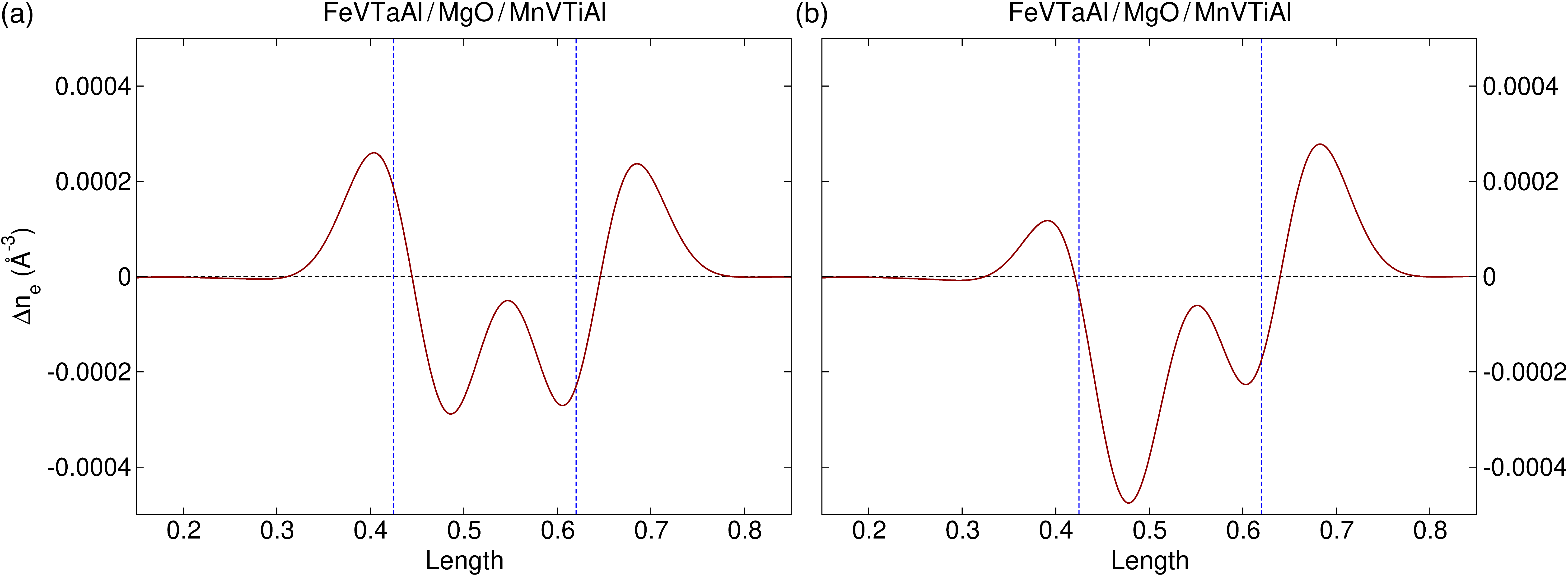}
    \caption{Electron difference density as function of the length of the FeVTaAl/MgO/MnVTiAl junction in (a) parallel and (b) anti-parallel orientation of the magnetization direction of the electrodes for 5ML of MgO at an applied bias of $-0.3$\,V. The dashed blue lines denote the interfaces with MgO and the dashed black line the zero line.}
    \label{fig10}
\end{figure}

For the FeVTaAl/MgO/MnTiVAl junction ((Fig.~\ref{fig4})) in parallel configuration, the picture differs slightly. Again, only majority carrier states are accessible in the vicinity of $E_F$. While FeVTaAl offers both $\Delta_1$ and $\Delta_5$ states, MnTiVAl only offers states of $\Delta_5$ character. Hence, only $\Delta_5$-like states will match and carry a current. Unfortunately, $\kappa_{\Delta_5} > \kappa_{\Delta_1}$ within the MgO barrier. Thus, by these qualitative arguments, there will be a spin-polarized current across the FeVTaAl/MgO/MnTiVAl junction, which is expected to be smaller than for the case of the FeVTiSi/MgO/CoFeVSb junction. 

For the anti-parallel configuration, the simplified discussion focusing on states around $\Gamma$ fails. It is evident that for no combination of minority$\rightarrow$majority or majority$\rightarrow$minority carrier states are simultaneously available in both channels close to $E_F$. We therefor extend our qualitative analysis to a more profound picture involving states off the $\Gamma$-point.

By analyzing the band structure in Fig.~\ref{fig5}, we first focus on the FeVTiSi/MgO/CoFeVSb junction in anti-parallel configuration. We note that states in the vicinity of $E_F$ along the $\Gamma-K$-line are available for the minority channel in FeVTiSi, as well as for the majority channel in CoFeVSb. Both channels only have states with $\Sigma_1$-symmetry representation in common. States with $\Sigma_1$-symmetry are compatible with states with $\Delta_1$-symmetry~\cite{Wigner:1936}. While we used the argument $\kappa_{\Delta_1} > \kappa_{\Delta_5}$ for the analysis around $\Gamma$, this simplified picture doesn't hold away from the Brillouin zone center. As can be seen from Figs.~\hyperref[fig6]{6\,(b)} and \hyperref[fig6]{6\,(c)}, the differences in the absolute values for $\kappa_{\Delta_1}$ and $\kappa_{\Delta_5}$ are most pronounced around $\Gamma$, but barely in other areas of the Brillouin zone. Especially along the $\overline{\Gamma}-\overline{X}$, i.e., a (001) projection of the $\Gamma-K$-line, it is $\kappa_{\Delta_1} \approx \kappa_{\Delta_5}$. 
As a consequence, we will have a current through the FeVTiSi/MgO/CoFeVSb junction in anti-parallel configuration dominated by $\Delta_1$-like states, which are moderately damped within the layers of MgO.

Fig.~\ref{fig7} depicts the band structures of the bulk ingredients of the FeVTaAl/MgO/MnTiVAl junction. Focusing on possible electronic transport in anti-parallel configuration, we decipher again states along the $\Gamma-K$-line as possible candidates to dictate transport. However, after close inspection, no states with common symmetry character can be found. While the minority carrier states of FeVTaAl do have $\Sigma_1$ character, the majority carrier states of MnTiVAl at $E_F$ do have $\Sigma_3$ and $\Sigma_4$ character. While $\Sigma_1$, $\Sigma_3$, and $\Sigma_4$ are decompositions of the $\Gamma_{15}$ representation, they are not compatible along the $\Gamma-K$-line. This symmetry mismatch leads to a vanishing current contribution in anti-parallel configuration for the FeVTaAl/MgO/MnTiVAl junction. Smaller contributions have to stem from heavily symmetry-mixed contributions off the high-symmetry lines and are accounted for by our full \textit{ab initio} calculations.

In the previous paragraphs, we discussed the symmetry character of the band structure of both MTJs and would like to comment now on the charge accumulation at the interfaces in the FeVTaAl/MgO/MnVTiAl junction.
In magnetic tunnel junctions, two different charge accumulations occur.
One in the electrode materials and another in the dielectric material, which arises from the capacitance of the MTJ.
The latter charge accumulation does not have an effect on the transport properties of the electrodes.
In the electrode materials, charge accumulation occurs due to the tunneling current and might influence their transport properties.
A similar charge accumulation can occur in the electrode materials due to a work function difference, which does not depend on a voltage or a current flowing through the device and also has an impact on the transport properties.
In the FeVTaAl/MgO/MnVTiAl junction, the work functions of the electrode materials differ by $200$\,meV, and thus just a small amount of charge is transferred from the half-metallic magnet to the spin-gapless semiconductor.
As one can see from Fig.~\ref{fig8}, in parallel and anti-parallel configuration, under zero bias, the electron density at the FeVTaAl/MgO interface increases slightly, and thus this interface gets negatively charged, while at the MnVTiAl/MgO interface, the electron density decreases and increases of the same amount a few layers away from the interface.

When a positive or negative bias voltage is applied to the device in the parallel orientation of the magnetization directions of the electrodes, the electron difference density at the FeVTaAl/MgO interface stays unchanged compared to zero bias (see Figs.~\hyperref[fig9]{9\,(a)} and \hyperref[fig10]{10\,(a)}), while in both cases the charge at the MnVTiAl/MgO interface decreases and increases within the tunnel barrier.

In the anti-parallel orientation of the magnetization direction of the electrodes, under forward bias ($+0.3$\,V), the behavior is similar to the parallel orientation, while under reverse bias ($-0.3$\,V), the largest changes occur. Here, the charge at the FeVTaAl/MgO interface and within the tunnel barrier decreases while it increases slightly at the MnVTiAl/MgO interface as well as a few layers away from the interface (see Figs.~\hyperref[fig9]{9\,(b)} and \hyperref[fig10]{10\,(b)}).

\bibliography{bibliography}